\documentclass{jfm}

\usepackage{algorithm2e}

\usepackage{amsmath}
\usepackage{amsfonts}

\usepackage{array}
\usepackage{booktabs}
    
\usepackage[usenames,dvipsnames]{xcolor}
\usepackage{tikz}
\usetikzlibrary{arrows,automata,shapes,calc}

\usepackage{pgfplots}

\makeatletter
\usepackage{natbib}

\usepackage{hyperref}
\hypersetup{
    colorlinks,
    linkcolor={red!70!black},
    citecolor={blue!70!black},
    urlcolor={blue!90!black}
}

\usepackage{graphicx}
\usepackage{epstopdf} 
\DeclareGraphicsExtensions{.pdf, .png}

\usepackage{cleveref}

\DeclareMathOperator{\tr}{tr}

\newcommand{\sDiv}{\nabla \cdot}

\newcommand{\sGrad}{\nabla}

\DeclareMathOperator{\sgn}{sgn}
\newcommand{\defgrad}{\mathbf{F}}

\newcommand{\cauchystress}{\boldsymbol{\sigma}}
\newcommand{\velgrad}{\mathbf{L}}
\newcommand{\symvelgrad}{\mathbf{D}}
\newcommand{\skwvelgrad}{\mathbf{W}}

\newcommand{\iden}{\mathbf{I}}

\usepackage{cleveref}

\newcolumntype{D}{ >{\centering\arraybackslash} m{4in} }
\newcolumntype{J}{ >{\centering\arraybackslash} m{3in} }
\newcolumntype{C}{ >{\centering\arraybackslash} m{2in} }
\newcolumntype{F}{ >{\centering\arraybackslash} m{1.5in} }
\newcolumntype{H}{ >{\centering\arraybackslash} m{1in} }

\newcolumntype{Y}{ >{\centering\arraybackslash} m{3.6in} }
\newcolumntype{Z}{ >{\centering\arraybackslash} m{2in} }

\usepackage[T1]{fontenc}
\usepackage[utf8]{inputenc}
\usepackage{lmodern}

\ifCUPmtlplainloaded \else
  \checkfont{eurm10}
  \iffontfound
    \IfFileExists{upmath.sty}
      {\typeout{^^JFound AMS Euler Roman fonts on the system,
                   using the 'upmath' package.^^J}%
       \usepackage{upmath}}
      {\typeout{^^JFound AMS Euler Roman fonts on the system, but you
                   dont seem to have the}%
       \typeout{'upmath' package installed. JFM.cls can take advantage
                 of these fonts,^^Jif you use 'upmath' package.^^J}%
      }
  \else
  \fi
\fi

\ifCUPmtlplainloaded \else
  \checkfont{msam10}
  \iffontfound
    \IfFileExists{amssymb.sty}
      {\typeout{^^JFound AMS Symbol fonts on the system, using the
                'amssymb' package.^^J}%
       \usepackage{amssymb}%
       \let\le=\leqslant  \let\leq=\leqslant
       \let\ge=\geqslant  \let\geq=\geqslant
      }{}
  \fi
\fi

\ifCUPmtlplainloaded \else
  \IfFileExists{amsbsy.sty}
    {\typeout{^^JFound the 'amsbsy' package on the system, using it.^^J}%
     \usepackage{amsbsy}}
    {\providecommand\boldsymbol[1]{\mbox{\boldmath $##1$}}}
\fi

\newcommand{\revised}[1]{{#1}}

\title{Continuum modeling and simulation of granular flows through their many phases}
\author[S. Dunatunga, K. Kamrin]{Sachith  Dunatunga$^1$ and  Ken  Kamrin$^1$}
\affiliation{$^1$Department of Mechanical Engineering, Massachusetts Institute of Technology, Cambridge, MA, 02139, USA}

\begin{document}
\maketitle

\begin{abstract}
We propose and numerically implement a constitutive framework for granular media that allows the material to traverse through its many common phases during the flow process.
When dense, the material is treated as a pressure sensitive elasto-viscoplastic solid obeying a yield criterion and a plastic flow rule given by the $\mu(I)$ inertial rheology of granular materials.
When the free volume exceeds a critical level, the material is deemed to separate and is treated as disconnected, stress-free media.
A Material Point Method (MPM) procedure is written for the simulation of this model and many demonstrations are provided in different geometries.
By using the MPM framework, extremely large strains and nonlinear deformations, which are common in granular flows, are representable.
The method is verified numerically and its physical predictions are validated against known results.
\end{abstract}

%
%
\section{Background}
Granular materials present several modeling challenges when considering a continuum approach. During dense flow, the material can be characterized as an elasto-viscoplastic material with a frictional yield criterion.  Extremely high levels of strain often occur, which challenge certain computational techniques, but the material can also behave as a solid, able to support shear loads in a static configuration.  Moreover, because dry grains do not support tension, their constitutive behavior changes from that of a dense plastic media to a gas-like disconnected state during extension, a dramatic switch that is difficult to represent in a unified modeling and numerical framework.

Several approaches have been used to simulate granular flow. One of the most accurate methods is the discrete element method (DEM), first described in \citet{cundall79}. While accurate, DEM solves the classical equations of motion on each grain individually, resulting in untenable computational expense over the large physical domains in many industrial and geological applications. A recent set of continuum rheological models for granular flow, such as the $\mu(I)$ relation in \citet{dacruz05} (later extended to 3D in \citet{jop06}) and the nonlocal extension in \citet{kamrin12}, offer a number of improvements over the commonly used rate-independent Drucker-Prager and Mohr-Coulomb models for problems with zones of dense, rapid flow (as is common in industrial settings) where rate-sensitivity is more pronounced and particle size-effects can play a role.  The incompressible Navier-Stokes solver Gerris has been used in \citet{staron12} and \citet{staron14} with the $\mu(I)$ relation, while the commercial finite-element software Abaqus was used in \citet{kamrin10} and appended with the nonlocal model in \citet{Henann13}. While both methods can yield good results in certain regimes, the fluid solvers have difficulties with extensional disconnection and truly static zones cannot be represented, while the finite-element method (FEM) has issues when mesh distortion becomes large. Fixes such as Arbitrary-Lagrangian Eulerian (ALE) re-meshing may cause loss of conservation guarantees and lead to severe errors in nontrivial constitutive relations.

The Material Point Method (MPM) combines the strengths of both the fluids solvers and the finite element method. First described in \citet{sulsky94}, MPM is a derivative of the fluid-implicit-particle (FLIP) method \citep{brackbill88}, which in turn is based on the particle-in-cell (PIC) method \citep{harlow64}. The key idea behind MPM is that the state of the simulation is contained in Lagrangian material points, while the equations of motion are solved on a background computational mesh in a manner similar to FEM. Importantly, since the state is saved at material points, the mesh can be reset at the beginning of each step. Using the updated Lagrangian formulation allows large deformations to accumulate on material points while the mesh generated at each step only deforms a small amount. Despite the use of this computational scratchpad, MPM falls into the ``meshless'' category of methods, as the connectivity between material points is not fixed and changes dynamically during the simulation. While the basic idea of MPM has been retained, variants exist which exhibit better behavior than the original formulation at the expense of more computation. More accurate integration methods were introduced with GIMP in \citet{bardenhagen04}. Recently, the Convected Particle Domain Interpolation (CPDI) method in \citet{sadeghirad11} and its improved variant CPDI2 in \citet{sadeghirad13} have been able to improve the accuracy of this integration even further at no extra computational cost over GIMP, provided that the deformation gradient is tracked on each material point. Alternatively, one can modify the background mesh as in Dual-Domain MPM from \citet{zhang11}, which can be used to suppress grid-crossing errors present when using the original formulation.

The natural ability of MPM to allow large deformations while retaining the ability of FEM to correctly handle elasto-static zones makes it attractive for modeling materials where both behaviors are present, such as in granular flow.
MPM practitioners have simulated individual grains with MPM \citep{bardenhagen00, bardenhagen01}, however we are interested in simulating the response as a bulk material.
\revised{ Recent work from \citet{abe13} and \citet{bandara15} shows that a coupled material point method works well for saturated soil problems involving large deformations, however other models are needed for dry granular materials when a gas phase is possible.
}
In \citet{andersen09}, simulation of a collapsing soil column is done with MPM, however no mention is made of possible tensile stress states.
Previous studies using MPM for studying granular flow during silo drainage \citep{wieckowski03, wieckowski04, wieckowski11} have prevented the material from entering a tensile stress state by assuming all extension can be reconciled as plastic dilation (with viscous regularization) per soil mechanics models \citep{schofield68}.
However, Reynolds dilation of this type is of a different fundamental nature from gas-like disconnection of grains, wherein the ability to support stress is essentially lost until the material collapses back to a dense state.
Our goal in this work is to produce a well-posed continuum model and corresponding simulation method that combines a realistic rate-dependent granular rheology with a disconnected state representation, permitting us to simulate in one setting a wide range of granular behaviors spanning several ``phases'': solid-like static behavior, plastic flow (up to very large strains), as well as separation and reconsolidation of the material.
Accounting for all these behaviors produces a robust method able to handle a large range of engineering problems, which we demonstrate through a variety of tests and examples.

%
%
\section{Theory}
We use the standard notation for continuum mechanics defined in \citet{gurtin10}, with the exception that the Cauchy stress is denoted by $\cauchystress$.
The spatial gradient and spatial divergence operators are given by $\sGrad$ and $\sDiv$ respectively. Material time derivatives are represented by an overdot.
\revised{ The trace of a tensor $\mathbf{A}$ is given by $\tr \mathbf{A}$, the transpose by $\mathbf{A}^T$, and the deviator by $\mathbf{A}_0 = \mathbf{A} - \frac{1}{3} (\tr \mathbf{A}) \iden$ (when in 3D). }


The equation for momentum balance is
\begin{align}
	\sDiv \cauchystress + \rho \mathbf{b} = \rho \mathbf{\dot{v}},
\end{align}
where $\rho$ is the density, $\mathbf{b}$ is the specific body force, and $\mathbf{\dot{v}}$ is the material rate of change of the velocity. We define the spatial velocity gradient as
\begin{align}
	\velgrad = \sGrad \mathbf{v},
\end{align}
which can be decomposed into the spin tensor $\skwvelgrad$ and strain-rate tensor $\symvelgrad$. The spin and strain-rate tensors are given by
\begin{align}
 \skwvelgrad &= \frac{1}{2}(\velgrad - \velgrad^T), \ \ \ \ \ \symvelgrad = \frac{1}{2}(\velgrad + \velgrad^T),
\end{align}
which are the skew and symmetric parts of $\velgrad$ respectively. The local form of mass balance can be written in terms of the density as
\begin{align}
	\dot{\rho} = -\rho \tr \velgrad.
	\label{eqn:local-density}
\end{align}

When the material is dense, our constitutive model resembles that of a Maxwell model whose damper obeys a Bingham-like rheology.  That is, we suppose a stiff elastic mechanism in series with a plastic flow mechanism, which permits us to maintain a well-defined stress even in sub-yield zones of material where plastic flow vanishes. To this aim, we adopt a rate-based, hypoelastic-plastic approach assuming an additive decomposition of the velocity gradient into elastic and plastic parts, $\velgrad=\velgrad^e+\velgrad^p$. The elastic and plastic parts can each be decomposed into their own spin and strain-rate tensors, denoted by $\symvelgrad^e$, $\symvelgrad^p$, and $\skwvelgrad^e$, $\skwvelgrad^p$ respectively. Although hypoelastic models do not explicitly utilize an elastic strain-energy potential, they more easily permit modeling  large inhomogeneous plastic deformations since reliance on the deformation gradient tensor is avoided.  Deleterious artifacts of hypoelasticity are minimized when elastic stretches are small, or more broadly, when only the eigenvalues of the elastic part of the reference stretch change.  These criteria are easily met in a dense granular flow of stiff grains.

We will adopt a plastic flow relation of the form
\begin{align}
\symvelgrad^p = \hat{\symvelgrad}^p(\cauchystress)
\end{align}
In the case of codirectionality and \revised{isochoric plastic deformation}, we can write
\begin{align}
\hat{\symvelgrad}^p(\cauchystress) = \frac{1}{\sqrt{2}}\dot{\bar{\gamma}}^p(\cauchystress) \frac{\cauchystress_0}{\| \cauchystress_0 \|},
\label{eqn:dp-codirectional}
\end{align}
where $\dot{\bar{\gamma}}^p$ is the equivalent plastic shear strain rate.  Reasonable codirectionality (with rotations of the principal axes only a handful of degrees apart) between $\symvelgrad^p$ and $\cauchystress_0$ in dense granular flow \revised{of stiff grains} has been observed in DEM simulations by \citet{dacruz05, silbert01, koval09}. The isochoric assumption assumes the  critical state of constant-volume flow is rapidly approached given the high strains and rates we will be modeling. Using this simplification, we can reduce our problem to that of determining $\dot{\bar{\gamma}}^p$ given a stress state. 

\revised{
For pressure, we use an equation of state given by
\begin{align}
    p(\rho) &= \begin{cases}
        0 & \text{ if } \rho < \rho_c \\
        \frac{K_c}{\rho} \left( \rho - \rho_c \right) & \text{ if } \rho \ge \rho_c.
    \end{cases}
    \label{eqn:eos}
\end{align}
The critical density $\rho_c$ is the density of the material when grains are beginning to lose contact (when the pressure has just become zero); below $\rho_c$ we model the material to be \emph{disconnected}.  The stress-free approximation when the density is below $\rho_c$ can be analyzed for accuracy in the context of kinetic theory. For details, please see \cref{sec:kinetic-theory}.
}

Since we will use a hypoelastic-plastic model for the dense flowing phase ($\rho > \rho_c$), we must choose an objective rate for the stress \revised{evolution equation, and the resulting form must be consistent with our equation-of-state for the pressure, Eq \eqref{eqn:eos}.}
We use the Jaumann rate here, which is defined by
\begin{align}
	\overset{\triangle}{\cauchystress} \equiv \dot{\cauchystress} - \skwvelgrad \cdot \cauchystress + \cauchystress \cdot \skwvelgrad.
\end{align}
Since we will be implementing an explicit scheme where time steps are sufficiently small, the numerical integration of the Jaumann rate maintains objectivity without the need for enhanced integration procedures such as in \citet{weber90} and \citet{rashid93}.
The constitutive relation is then written as a relation between the  elastic strain-rate, $\symvelgrad^e=\symvelgrad-\symvelgrad^p$, and the stress-rate, i.e.
\begin{align}\label{Jaumann}
\overset{\triangle}{\cauchystress} = \mathbb{C} \colon \symvelgrad^e \equiv \frac{E}{1+\nu}\left((\symvelgrad - \symvelgrad^p)+\frac{\nu}{1-2\nu}\tr (\symvelgrad - \symvelgrad^p)\iden\right)
\end{align}
where $\mathbb{C}$ is the fourth-order elastic stiffness tensor, which depends on Poisson's ratio $\nu$ and Young's modulus $E = 3 K_c (1 - 2\nu)$.  \revised{More explanation on the connection between Eqs \eqref{eqn:eos} and \eqref{Jaumann} is in  \cref{sec:kinetic-theory}.}

We define the Drucker-Prager friction coefficient $\mu$ as $\bar{\tau} / p$, where in 3D, the quantities $\bar{\tau}$ and $p$ are in turn defined from the stress tensor via
\begin{align}
\bar{\tau} &= \sqrt{\frac{1}{2}(\cauchystress_0 \colon \cauchystress_0)}, \qquad p = -\frac{1}{3} \tr \cauchystress,
\end{align}
and $:$ denotes the tensorial contraction (defined for tensors $\mathbf{A}$ and $\mathbf{B}$ as $\mathbf{A} \colon \mathbf{B} = \sum_{i}\sum_{j} A_{ij} B_{ij}$). Although the model we are using is a fully 3D formulation, our examples are done in plane strain (that is, $\symvelgrad_{zz} = 0$ but $\cauchystress_{zz} \ne 0$).

The local model presented in \citet{jop06}, which has been validated experimentally in that work and  numerically in \citet{dacruz05}, relates the inertial number $I$  given by
\begin{align}
	I = \dot{\bar{\gamma}}^p \frac{\sqrt{d^2 \rho_s}}{\sqrt{p}}
	\label{eqn:I}
\end{align}
to the friction coefficient $\mu$ through the equation
\begin{align}
	\mu=\mu(I) = \mu_s + \frac{\mu_2 - \mu_s}{I_0 / I + 1} \ \ \text{if} \ \ I>0, \ \ \text{and} \ \  \mu\leq\mu_s \ \  \text{if} \ \ I= 0.
	\label{eqn:jop-muI}
\end{align}
 Here $I_0$ is a material constant, $\rho_s$ is the density of solid grains, and $d$ is the mean particle size (which only serves to scale the rate sensitivity, i.e. it is not an intrinsic length scale in the model).

\revised{Although the simplifications made in Eqs \eqref{eqn:eos} and \eqref{eqn:dp-codirectional} neglect flow-induced dilation effects in dense shear flow, the above works have also found a $\Phi(I)$ relation, which relates the packing fraction to the inertial number in steady flows.
More discussion of this, and how the $\Phi(I)$ relation can be used post-facto to correct the density field prediction in our computed results can be found in \cref{sec:kinetic-theory}.}

Since $\mu$ is related to the stress through $\bar{\tau}$ and $p$, and $I$ is related to $\dot{\bar{\gamma}}^p$, the above is a closed rheological relation. We can rewrite this into a rate-dependent form for the equivalent shear stress, given by
\begin{align}
\bar{\tau}=\bar{\tau}(p, \dot{\bar{\gamma}}^p) = p \left(\mu_s + \frac{\mu_2 - \mu_s}{\xi \sqrt{p}/\dot{\bar{\gamma}}^p + 1}\right) \ \ \text{if} \ \ 
\dot{\bar{\gamma}}^p>0,\ \ \text{and} \ \  \bar{\tau}\leq p\mu_s \ \ \text{if} \ \ \dot{\bar{\gamma}}^p=0.
\label{eqn:rate-form-tau-gammadotp}
\end{align}
where the constant $\xi = I_0/\sqrt{d^2 \rho_s}$ scales the rate-sensitivity.
We observe that $\mu_s$ is a static friction coefficient; no plastic flow occurs when $\mu<\mu_s$.  Our plasticity model is defined solely through equations \ref{eqn:dp-codirectional} and \ref{eqn:jop-muI};  plastic transients are neglected by assuming the shear required to reach a ``critical-state'' is negligible compared to the expected deformation levels. 


Combining \revised{the behavior in the solid, flowing, and gas phases},  if the material is below the critical density, the material is \emph{disconnected} and treated as stress-free. Otherwise, the material is \emph{dense} and has a positive pressure, so we are able to use the elasto-viscoplastic relation. Altogether, we write
\begin{align}
\cauchystress &= \mathbf{0} \qquad\text{ if } \rho < \rho_c
\label{eqn:constitutive-stress-free}
\\
\overset{\triangle}{\cauchystress} &= \mathbb{C}\colon(\symvelgrad - \hat{\symvelgrad}^p(\cauchystress)) \qquad\text{ otherwise },
\label{eqn:constitutive-cases}
\end{align}
where $\hat{\symvelgrad}^p$ is obtained from equations \eqref{eqn:dp-codirectional} and \eqref{eqn:rate-form-tau-gammadotp}.  Since $\hat{\symvelgrad}^p(\cauchystress)$ vanishes when $\mu<\mu_s$ giving an elastic state without plastic flow, we recognize the above rule as essentially a three-phase constitutive relation that can transition between elastic solid-like behavior, viscoplastic fluid-like flow, and dilute disconnected granular behavior. 

%
%
\section{Algorithm}
\subsection{Material Point Method Implementation}
Numerically implementing our proposed system requires a robust framework.
We propose to use MPM, a topic of active research, which has many variants.
\revised{The basic idea of MPM is to store the mechanical information, such as stress, momentum, and mass, on a set of  Lagrangian material points.  As visualized in the left panel of Fig \ref{fig:mpm-update-order}, a few steps take place during a time-step, with the end goal of updating the positions of the points and the quantities stored on them.  First, a background finite-element mesh is introduced (we choose a simple cartesian mesh), and the mechanical quantities stored on the points are projected onto the nodes of the mesh.  The mesh now has the needed information to conduct a single finite-element update step;  the constitutive relation and the equations of motion are used to move the mesh itself and update the quantities stored on the nodes.  The solution from the finite-element mesh is projected back onto the material points, which updates their internal state and moves them.  Once the material points have been updated and moved, the distorted finite-element mesh is destroyed; in the next step, a new cartesian mesh is introduced. In this manner, mesh entanglement issues are avoided.  In MPM, the material points are the persistent Lagrangian domain of the scheme and the mesh appears mid-step, temporarily, only to organize the update of the points' motion and state.
}

For the projection steps, we can choose various particle characteristic functions -- $\chi_{p}$ in the notation of \citet{bardenhagen04}, or more generally can find different methods to perform the integration as in \citet{sadeghirad11} and \citet{zhang11}.
In moving the background mesh, we can use an explicit scheme as in the original MPM formulation, or we can use an implicit scheme which is similar to that employed in finite element methods, as in \citet{guilkey03}.
All of these variants maintain the original spirit of MPM, in that information is projected from mobile state carriers to a background mesh, but for clarity we detail our implementation below.
The diagram of information flow for our algorithm is shown in figure \ref{fig:mpm-update-order}.
Typically, the particle volume does not directly enter the stress update step, however this is a key feature of the current work. We use the same notation as in \citet{bardenhagen04}. In our case, we use scaled delta functions for the particle characteristic functions as in the original MPM formulation from \citet{sulsky94}. Although more accurate schemes exist, they usually involve spreading the influence of the material point (GIMP, DDMP) or rely on the deformation gradient tensor at a material point (CPDI). 



First order elements are used for the background mesh, as second order elements have issues unique to MPM as noted in \citet{andersen10}. For a regular Cartesian mesh with spacing $\Delta x$, the shape functions and gradients in 1D for node $i$ are given by
\begin{align}
	S_i(x) &= \max{(0, (1 - \frac{1}{\Delta x}|x_i - x|))} \\
	\nabla S_i(x) &= \begin{cases}
	\frac{1}{\Delta x} \sgn{(x_i - x)}  & \mbox{\text{if }} |x_i - x| \le \Delta x \\
		0 & \mbox{\text{otherwise}}
	\end{cases}
\end{align}
respectively, where $x_i$ is the position of the node. Products of these functions can be used to generate the shape functions for higher dimensions. When combined with the scaled delta functions, the 1D mapping functions and gradient mapping functions for material point $p$ are given simply by
\begin{equation}
	S_{ip} = S_i(x_p), \quad \nabla S_{ip} = \nabla S_i(x_p),
\end{equation}
where $x_p$ is the position of the material point.

At the beginning of an MPM step, we produce the nodal quantities from material point quantities via the sums
\begin{align}
m^n_i &= \sum_{p} S_{ip} m_p, \ \ \ \  \ (m\mathbf{v})^n_i = \sum_{p} S_{ip} m_p \mathbf{v}^n_p, \ \ \ \  \ \mathbf{b}^n_i = \sum_{p} S_{ip} m_p \mathbf{b}^n_p, \ \ \ \ \ \mathbf{f}^n_i = \sum_{p} -v_p \cauchystress^n_p \cdot \nabla S_{ip},
\end{align}
where $m_p$ is the (constant) particle mass, and $\mathbf{v}^n_p$, $\mathbf{b}^n_p$,  and $\cauchystress^n_p$ are the particle velocity, specific body force, and Cauchy stress respectively at the beginning of the time step. The nodal quantities, indexed by $i$, are given in order as the mass, momentum, body force, and internal forces at the beginning of the time step.
The change in momentum on the nodes is then given by
\begin{align}
\dot{(m\mathbf{v})}^n_i = \mathbf{b}^n_i + \mathbf{f}^n_i.
\end{align}
Using a forward Euler integration, we write the nodal momentum as
\begin{align}
(m\mathbf{v})_i^{n+1} = (m\mathbf{v})_i^n + \Delta t \dot{(m\mathbf{v})}^n_i
\end{align}
Since the nodal mass remains constant within a time step, we can then write the nodal acceleration and velocity respectively by
\begin{align}
\dot{\mathbf{v}}_i^{n+1} = \frac{\dot{(m\mathbf{v})}_i^n}{m_i}, \qquad
\mathbf{v}_i^{n+1} = \frac{(m\mathbf{v})_i^{n+1}}{m_i}.
\end{align}
These nodal quantities are then mapped back onto the material points via the equations
\begin{align}
\mathbf{v}^{n+1}_p &= \mathbf{v}^{n}_p + \Delta t \sum_{i} S_{ip} \mathbf{\dot{v}}^{n+1}_i, \ \ \ \ \ \velgrad^{n+1}_p = \sum_{i} \mathbf{v}^{n+1}_i \otimes \nabla S_{ip}, \ \ \ \ \ \mathbf{x}^{n+1}_p = \mathbf{x}^{n}_p + \Delta t \sum_{i} S_{ip} \mathbf{v}^{n+1}_i.
\end{align}
Although the sums in these equations are over all the nodes, in practice the particles only contribute to a very small subset of nodes and vice-versa. With our implementation, each particle exists at most in one quadrilateral, so there are at most 4 nonzero values of $S_{ip}$ per particle. Other variants of MPM may have more possible non-zeroes (up to 16 with both CPDI and GIMP on a Cartesian mesh), but for realistic problems the matrix of mapping functions is never dense.

We use the particle velocity gradient to update the particle stress through the constitutive update, which will be detailed in a later section. Others \citep{sadeghirad11,nair12} have used the deformation gradient $\defgrad$ to update the particle stress, e.g. in a hyperelastic model, however as stated before we wish to avoid reliance on $\defgrad$ due to the large inhomogeneous deformations.

For constitutive purposes it is also important to track particle-level density, which we achieve by updating the local volume of each material point.  For a constant velocity gradient $\velgrad$, the analytical solution for \eqref{eqn:local-density} is given by $
\rho(t) = \rho_0 \exp(-t \tr \velgrad)$ where $\rho_0 = \rho(0)$ is the initial density.  Since the mass of a material point remains constant, the density $\rho_p$ is inversely proportional to $v_p$, the volume of the material point.  Substituting this result, and considering the velocity gradient to be a constant over one time step, we can obtain a rule for evolving the volume of a material point, which in turn gives the material point density:
\begin{align}
v^{n+1}_p = v^{n}_p \exp (\Delta t  \tr \velgrad^n), \ \ \Longrightarrow \ \ \rho^{n+1}_p = \frac{m_p}{v^{n+1}_p}.
\end{align}

\subsection{Stress Update}
The last quantity that is updated during an MPM step is the stress, $\cauchystress^{n+1}_p$, which we calculate using equations  \eqref{eqn:constitutive-stress-free}  and \eqref{eqn:constitutive-cases}, as discretized in terms of $\rho^{n+1}_p $, $\velgrad^{n+1}_p$, and $\cauchystress^n_p$.  For the rest of this section, we neglect the $p$ subscript for ease.

By equation \eqref{eqn:constitutive-stress-free}, in the simplest case where $\rho^{n+1}$ is below a critical value $\rho_c$, we set $\cauchystress^{n+1}= \mathbf{0}$. For all other cases, we proceed with an analog of the elastic trial step commonly used in numerical plasticity. In this manner, we follow the method of \citet{kamrin10}, with modifications for using simple linear elasticity in hypoelastic-plastic form instead of the more complex Jiang-Liu model \citep{jiang03} and a hyperelastic framework.  The ``trial stress'' at a material point is calculated from
\begin{align}
	\cauchystress^{tr} = \cauchystress^n + \Delta t (\mathbb{C} \colon \symvelgrad^{n+1} + \skwvelgrad^{n+1} \cdot \cauchystress^n - \cauchystress^n \cdot \skwvelgrad^{n+1}).
\label{eqn:trial-stress}
\end{align}
This allows us to resolve the other portion of the conditional in equation \eqref{eqn:constitutive-stress-free} in a simple manner. We first define $\Delta p = p^{tr} - p^{n}$. Expanding equation \eqref{eqn:trial-stress} and taking the trace yields $\Delta p = -K \Delta t \tr \symvelgrad$, where $K \equiv \frac{E}{3(1-2\nu)}$ is the bulk modulus. Upon examination of equation \eqref{eqn:local-density}, we note that both $\dot \rho$ and $\Delta p$ are proportional to $\tr \velgrad$, since $\tr \velgrad = \tr \symvelgrad$. Since the pressure at the critical density is zero, if the material is currently at the critical density, the pressure at the end of the trial step is $\Delta p$.  Hence, if the trial pressure is negative, the density is decreasing through $\rho_c$, so $\cauchystress^{n+1}= \mathbf{0}$ by \eqref{eqn:constitutive-stress-free}.

If the material point passes both checks, it is in the correct regime to use equation \eqref{eqn:constitutive-cases}.  Defining
\begin{equation}
S_{0} = \mu_s p^{tr}, \ \ \ \  S_{2} = \mu_2 p^{tr}, \ \ \ \  \alpha = \xi G \Delta t \sqrt{p^{tr}}, \ \ \ \  B = S_2 + \bar{\tau}^{tr} + \alpha,  \ \ \  \ H = S_2 \bar{\tau}^{tr} + S_0, \alpha
\end{equation}
the material point is in the elastic regime (no plastic flow) if $\bar{\tau}^{tr} \leq S_0$.  In this case we set $\cauchystress^{n+1}= \cauchystress^{tr}$.  

If $\bar{\tau}^{tr} > S_0$, then plastic flow must occur and we must determine the value $\dot{\bar{\gamma}}^p$.
For stability reasons we utilize an implicit constitutive update, which means the equivalent plastic shear flow rate $\dot{\bar{\gamma}}^p$ is consistent with the value of $\mu$ at the end of the step, i.e. $\mu(I) = \mu^{n+1}$, where $I = \xi^{-1} \dot{\bar{\gamma}}^p / \sqrt{p^{n+1}}$. Since we assume plastic incompressibility, we know that the trial pressure $p^{tr}$, calculated assuming the entire deformation was elastic, is the final pressure $p^{n+1}$. To find the relation for the equivalent shear stress $\bar{\tau}^{n+1}$, we first note that we can write the stress as
\begin{align}\label{eqn:sig_update}
\cauchystress^{n+1} = \cauchystress^{tr} - \Delta t \mathbb{C} \colon \symvelgrad^p.
\end{align}

Define the shear modulus $G$ as $\frac{E}{2(1+\nu)}$. Noting that $\mathbb{C} \colon \symvelgrad^p = 2G \symvelgrad^p$, and that $\cauchystress^{tr}_0$ and $\cauchystress^{n+1}_0$ are codirectional, we can write the equation for $\bar{\tau}^{n+1}$ as
\begin{align}
 \bar{\tau}^{n+1} = \bar{\tau}^{tr} - G \Delta t \dot{\bar{\gamma}}^p.
\end{align}
We can rewrite equation  \eqref{eqn:rate-form-tau-gammadotp} simply as the quadratic
\begin{align}\label{eqn:quadratic}
(\bar{\tau}^{n+1})^2 - B \bar{\tau}^{n+1} + H = 0,
\end{align}
where $\bar{\tau}^{n+1}$ is the only unknown. Solving for $\bar{\tau}^{n+1}$ gives us two roots; only the negative root has physical meaning, as the positive root implies a negative equivalent plastic shear strain rate. 

\begin{figure}
\begin{tabular}{Y !{\vrule width 2pt} Z}
\includegraphics[scale=1]{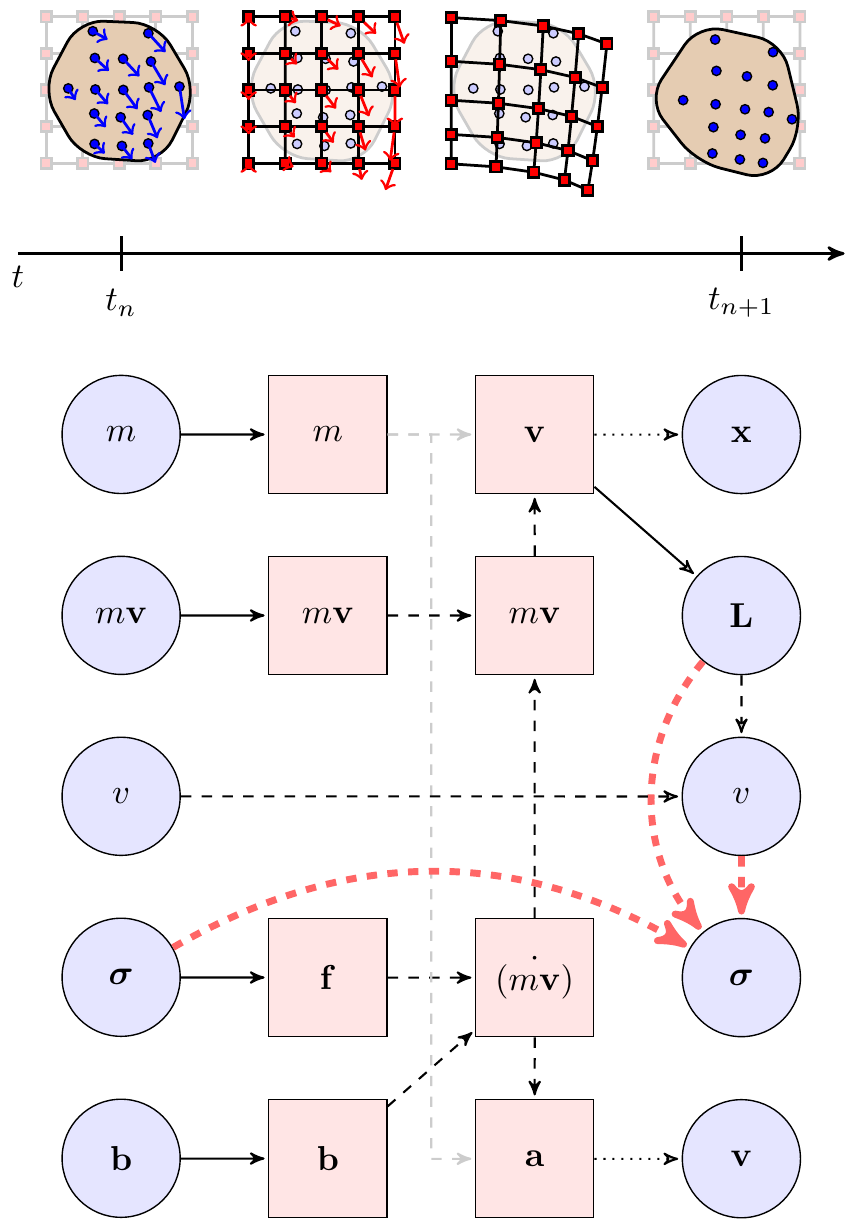} & \includegraphics[scale=1]{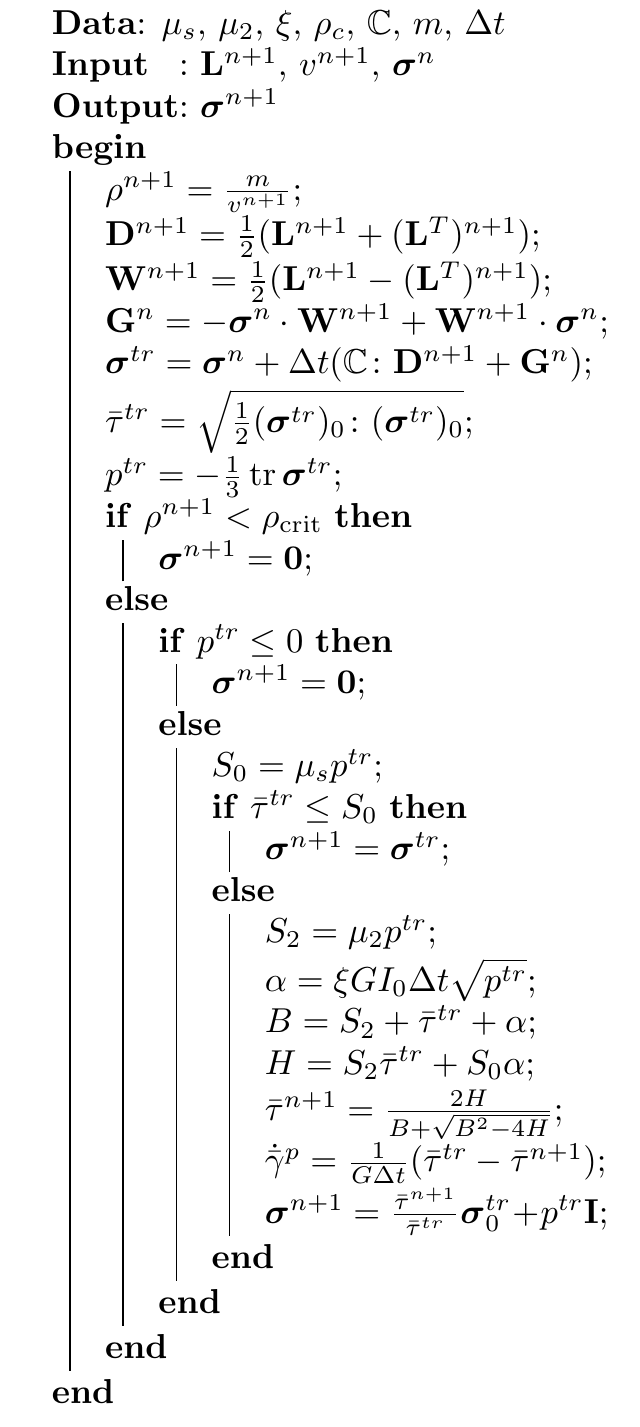}
\end{tabular}
\caption{Left panel: The steps needed to perform an explicit MPM step. Reading from left to right shows the order in which properties are modified within a time step. Material point properties are drawn with a circle, while nodal properties are drawn with squares. The diagram above the timeline shows the collective behavior of many material points and nodes, and is synchronized with the lower part of the diagram. Solid lines are projections using either the shape functions or gradient of shape functions, dashed lines are update operations, and dotted lines are a combination of projection and update. Note that multiple sources are needed for the calculation of some quantities. The thick dashed lines are the components of the stress update algorithm. (Diagram inspired by one in \citet{buzzi08}). Right panel: The stress update algorithm. All operations take place on the material point level, and each material point can be updated independently of any other material point, so subscripts have been dropped. 
}
\label{fig:mpm-update-order}
\end{figure}
There are some numerical subtleties in carrying out the described update of $\bar{\tau}$ and the subsequent update of $\cauchystress$.  The numerical stiffness of \eqref{eqn:quadratic}  near $\mu$ = $\mu_2$ is high.  Moreover in the flowing state the plastic strain-rate tensor $\symvelgrad^p$ may be very close to the total strain-rate tensor $\symvelgrad$. Hence, the subtraction of the two tensors in \eqref{eqn:sig_update} to obtain $\cauchystress^{n+1}$ may be inaccurate due to the loss of many significant digits.  

We remedy these issues as follows. Noting that $B$ is always positive, it is numerically advantageous to simplify the output of the quadratic equation when we solve for $\bar{\tau}^{n+1}$.  Specifically, we use
\begin{align}
	\bar{\tau}^{n+1} = \frac{2 H}{B + \sqrt{B^2 - 4 H}}.
\end{align}
This avoids the possible cancellation error of $B$ and $\sqrt{B^2 - 4H}$. Although the discriminant may still prove to be a source of error, the numerical fix in that case is not quite as straightforward \citep{kahan04}. Regarding $\cauchystress^{n+1}$, we simply use the answer obtained from $\bar{\tau}^{n+1}$ to scale the deviator of the trial stress. That is, we compute the particle stress at the end of the step as
\begin{align}
	\cauchystress^{n+1} = \frac{\bar{\tau}^{n+1}}{\bar{\tau}^{tr}}\cauchystress^{tr}_0 + p^{tr} \iden.
\end{align}
In practice we have found that these two fixes significantly improved the quality and stability of the simulation compared to the naive implementation. The entire stress update procedure is shown in the right panel of figure \ref{fig:mpm-update-order}. This stress update procedure is particularly amenable to parallel computation, as each particle can be updated independently with no synchronization required. We took advantage of this in our implementation. Verification of our implementation of the numerical scheme is carried out in \cref{app:verification}.

%
%
\section{Results}

Before showing a number of tests and demonstrations, we provide a brief discussion on how certain quantities are visualized.  There are several ways to plot stress-related variables in MPM; we can plot the material point markers themselves as having a stress value, or we can use the background mesh to smooth the stress results. High frequency errors are seen in plotting raw material point stress data, and may lead to an anomaly in MPM termed kinematic locking by \citet{mast12}. Techniques to remove this behavior are developed in the same work, however they do not consistently resolve the issue, and we do not implement those methods here. While in the material point case we are plotting exactly what is stored at specific locations, using the grid is attractive because it allows us to plot what the grid interprets when solving the equations of motion. We generally follow the recommendations in \citet{andersen13}, who present a more through discussion on post-processing, and utilize the method which maps material point values to nodes and then projects them back onto the material points for plotting purposes. 
Briefly, we use the mapping functions $S_{ip}$ and the material point mass to construct a mass-weighted nodal stress via
\begin{align}
(m\cauchystress)_i = \sum_{p} S_{ip} m_p \cauchystress_p.
\end{align} The stress $\tilde{\cauchystress}$ is then computed at the nodes as
\begin{align}
\tilde{\cauchystress}_i = \frac{(m\cauchystress)_i}{m_i},
\end{align}
which is projected back onto the material points using
\begin{align}
\tilde{\cauchystress}_p = \sum_{i} S_{ip} \tilde{\cauchystress}_i.
\end{align}
We reiterate that this procedure is purely a post-processing operation for visualization purposes. 
This is only applied to stress related variables, such as the pressure, and not to deformation related variables such as equivalent plastic shear strain rate.

\revised{
Several geometries are simulated using the method: granular column collapse, drainage of a silo, flow down an incline plane, and the impact of a granular slug.
The column collapse problem has been studied extensively in the literature and provides a good benchmark against both experiments and other numerical methods.
The silo drainage problem has been studied experimentally and with the discrete element method (and recently with the Gerris Navier-Stokes solver), but is typically outside the range of most continuum methods.
Others have used MPM \citep{wieckowski03} in this context with different material models, however we will provide a more quantitative test with the Beverloo correlation.
Flow down an inclined plane is a good test for the verification of the $\mu(I)$ rheology in the code, as we can analytically determine the flow profile as well as average flow rate as a function of tilt angle.
Finally, the impact of the granular slug, while non-quantitative, demonstrates the ability of the method to model multiple phases of granular flow with physically realistic results.
}

Although the initial configurations have different geometries, the material points all begin from a stress-free state (that is, $\cauchystress^{n=0}_p = \mathbf{0}$). We also apply gravity in a consistent manner across all simulations (excepting the granular slug example); gravity is ramped linearly from zero to its final value of 9.8 $\mathrm{m/s}^{2}$ at 0.1s and then held at this constant value for the rest of the simulation. When we describe boundary conditions, fully-rough walls are those in which a particle at the surface can neither move laterally along the boundary nor go through the boundary. In MPM, this is implemented as fixing all components of momentum along the boundary nodes to zero. Momentum updates are also supressed on these nodes, and the effect of both operations is that the mesh displacement at these nodes is zero. Frictionless walls do not allow particles to go through a boundary, but do allow unrestrained motion laterally along the boundary. Similar to the fully rough condition, in MPM we set the component of momentum perpendicular to the boundary to zero (and suppress the momentum update of this component as well), but we do not modify the parallel component of momentum. For our simulations, the symmetric boundary condition reduces to the same set of conditions as the frictionless wall. Periodic boundary conditions are implemented by mapping two distinct physical nodes (slave nodes) to the same logical node (master node), although the set of shape functions and shape function gradients remain distinct due to different physical positions. Thus accessing or modifying either of the slave nodes will access or modify data on the same master node (again, excepting the shape functions and shape function gradients). This gives the slave nodes the same state as each other (due to the shared underlying master node) and automatically transfers information across the boundary. The modulus operator is applied to the material point position to wrap material points around the domain (i.e. materials points which leave one side of the periodic boundary will re-emerge on the other size).

The values of $\nu$, $\xi$, $\mu_s$, and $\mu_2$ were kept constant for all the simulations and given by the values in table \ref{tbl:material-params}.
The plastic flow parameters correspond approximately to the glass bead parameters in \citet{jop06}.
The solid grain density throughout is set to $\rho_s = 2450$ kg $\mathrm{m}^{-3}$, \revised{which is a typical value for glass}. In line with 3D bead packings, we assume the material has a uniform initial packing fraction of $\Phi_0 \approx 60\%$, implying the initial density is \revised{$\Phi_0\rho_s = \rho_c = 1500$ kg $\mathrm{m}^{-3}$}. \revised{The specific value chosen for the critical packing fraction $\Phi_c = \rho_c / \rho_s$ does not affect the results shown within a reasonable range for packing fraction (within a few percent).} The granular slug starts with a slightly lower density of $1200$ kg $\mathrm{m}^{-3}$ at a packing fraction of $\Phi \approx 49\%$, however the critical density remains the same as in the other simulations. Additional simulation parameters are presented in table \ref{tbl:simulation-params}.

\begin{table}
\centering
\caption{Common material parameters.}
\begin{tabular}{r | l l l l l}
Parameter & $E$ & $\nu$ & $\xi$ & $\mu_s$ & $\mu_2$ \\
\midrule
Value & 1 GPa & 0.3 & $1.1233 \sqrt{\frac{\mathrm{m}}{\mathrm{kg}}}$ & 0.3819 & 0.6435 \\
\end{tabular}
\label{tbl:material-params}

\end{table}
\begin{table}
	\centering
	\caption{Simulation parameters.}
	\begin{tabular}{r l l l l}
		Parameter & Column Collapse & Silo and Hourglass & Incline & Slug\\
		\midrule
		$\Delta x$ & 0.0125/0.00625 m & 0.01 m & 0.005 m & 0.0083 m \\
		$\Delta t$ & $3 \times 10^{-6}$ s & $3 \times 10^{-6}$ s & $3 \times 10^{-6}$ s & $3 \times 10^{-6}$ s 
	\end{tabular}
	\label{tbl:simulation-params}
\end{table}

\subsection{Column Collapse}
\begin{figure}
	\centering
	\begin{tabular}{r F F F}
		Time & $a = 2.0$ & $a = 1.0$ & $a = 0.5$ \\
		\midrule
		0s & \includegraphics[scale=0.75]{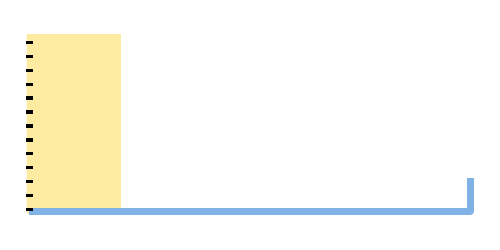} & \includegraphics[scale=0.75]{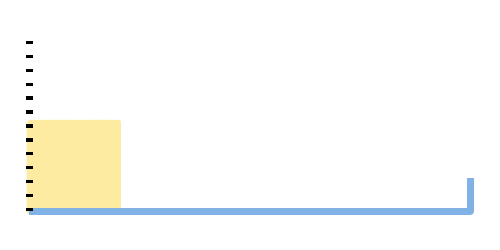} & \includegraphics[scale=0.75]{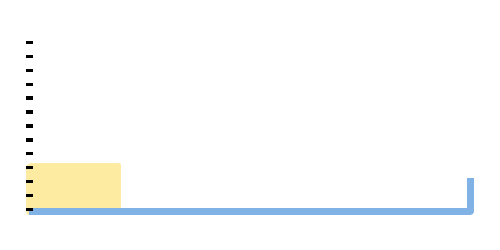} \\
		0.5s & \includegraphics[scale=0.75]{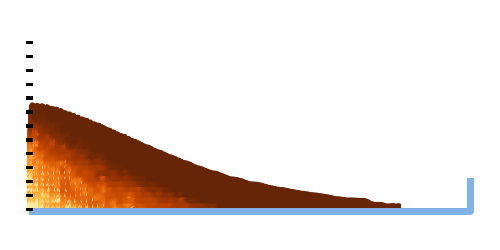} & \includegraphics[scale=0.75]{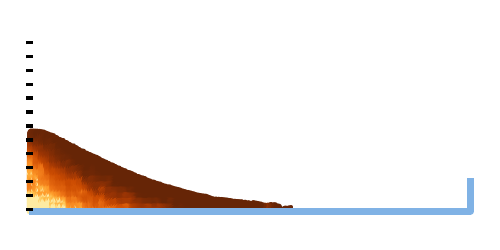} & \includegraphics[scale=0.75]{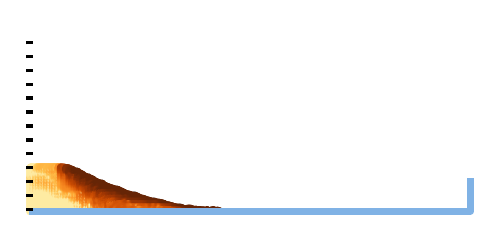} \\
		2.0s & \includegraphics[scale=0.75]{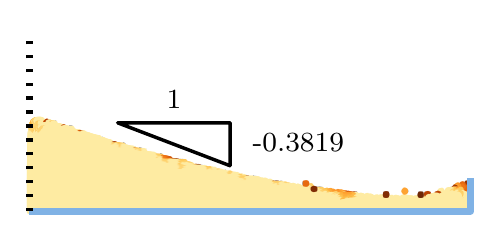} & \includegraphics[scale=0.75]{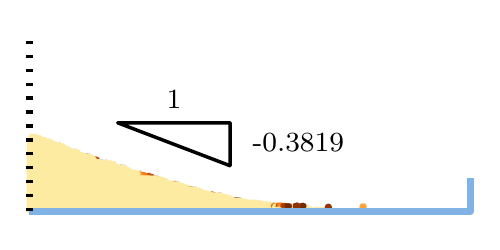} & \includegraphics[scale=0.75]{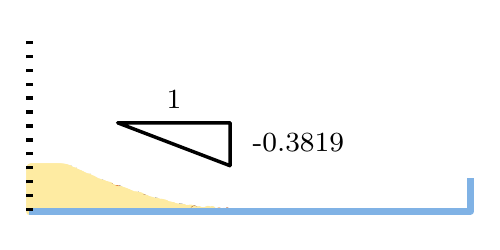} \\
		& \multicolumn{3}{c}{\includegraphics[scale=1.0]{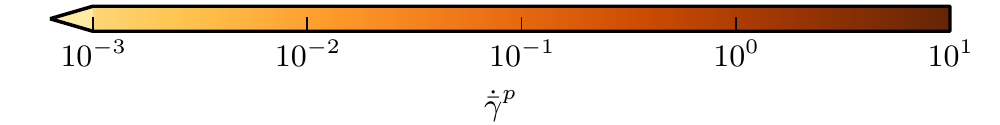}} \\
		2.0s & \includegraphics[scale=0.75]{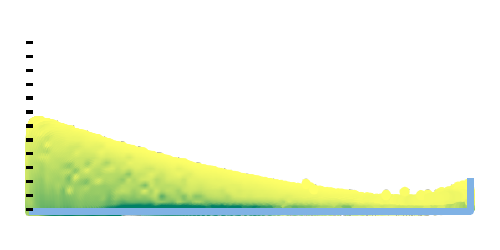} & \includegraphics[scale=0.75]{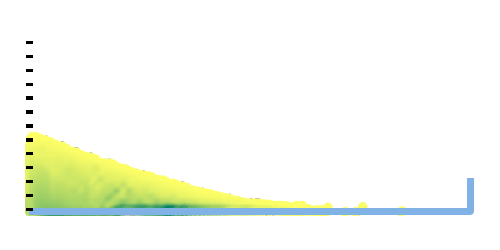} & \includegraphics[scale=0.75]{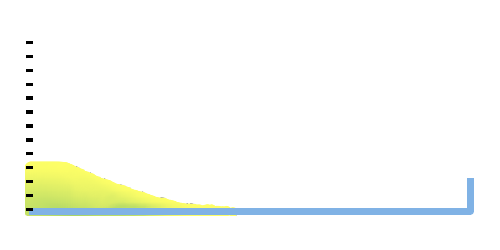} \\
		& \multicolumn{3}{c}{\includegraphics[scale=1.0]{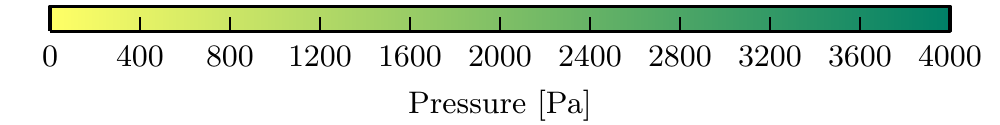}} \\
	\end{tabular}
	\caption{
	Column collapse: All particles shown are the material point markers (not grains). The half-height $w$ of each column is 20cm, while the heights are 40cm, 20cm, and 10cm from left to right.   We see that the resultant heaps follow the correct trend of higher aspect ratios leading to flatter heaps, and that the repose angle is less than the value implied by the static friction coefficient. The top three rows of images show the equivalent plastic shear-rate as the simulation progresses, while the bottom row of images shows the pressure distribution in the heap at the end of the simulation.}
	\label{fig:heap-images}
\end{figure}

We first tested our method on the problem of a granular column collapsing into a heap. Due to the static yield criterion, the material must settle at or below the maximum repose angle of $\arctan (\mu_s)$. In our case, this is approximately 20.9 degrees. Columns with a larger aspect ratio of height to width will tend to flatten out more due to macro-inertial effects, which has been observed in discrete element simulations by \citet{lacaze09} and experimentally by \citet{lube04}. This test is important as it indicates the dense $\mu(I)$ portion of the code is behaving as expected, and indeed we replicate the correct trend as shown in figure \ref{fig:heap-images}. Consistent with other literature on the step collapse, we define the aspect ratio $a$ as the height over the half-width of the heap. The top row shows the initial configuration of three examples having aspect ratio 2.0, 1.0 and 0.5 from left to right. The half-width is held constant at 0.20m (only the right half of the symmetric problem is computed and shown). The bottom boundary condition is a fully rough surface, while the symmetric boundary on the left side of the figure allows particles to slide along the vertical axis freely but prevents motion across the boundary. The right `lip' is also a frictionless wall. The variable plotted is ${\dot{\bar{\gamma}}^p}$. At 0.5s, the material has deformed significantly, as indicated in the second row. At 2.0s, the heaps have largely reached their static configurations, indicated by the small strain-rates; note that in all cases the repose angle is less than or equal to $\arctan (\mu_s)$.  The two larger column-collapse simulations used a mesh size $\Delta x$ of 0.0125m, while the smallest used a mesh size $\Delta x$ of 0.00625m; the time step $\Delta t$ of $3 \times 10^{-6}$s remained unchanged as it is stable for both mesh sizes. The elements are square, so $\Delta x = \Delta y$. We used 16 material points per occupied element for the initial distribution. For all the other simulations in this paper, we began with 4 material points per occupied element. 

Most of the heap becomes static in a short time, however, in the true continuum limit, it can take an unboundedly long time for the material at the free-surface to stop as the surface angle approaches $\arctan (\mu_s)$, due to the $\mu(I)$ rate-sensitivity function. The finite number of material points may allow it to reach the angle at a finite time, but even when the displacements appear to be small we may still observe a thin, slow-moving layer of particles at the free surface.

The fourth row of figure \ref{fig:heap-images} shows the pressure distribution in each of the heaps at 2.0s. Although \citet{geng01} show the pressure distribution depends on the preparation method, we see that using this particular numerical protocol the pressure dip found experimentally in \citet{brockbank97} is replicated with MPM. The aforementioned procedure was used to plot the pressure, and simply using the material point values `as is' results in significantly noisier data.

The simulations took approximately 20 to 30 minutes each on an Intel i5-4200U using 4 threads, taking more time for the larger heap and less for the smaller, due to the varying number of material points. We ran the simulations to 5 seconds for each heap, but found only 2 seconds were necessary before the simulation became static almost everywhere.

\revised{
Additional tests were performed to compare the aspect ratio against the run-out length, which is the half-width of the final configuration minus the half-width of the initial configuration.
This quantity is nondimensionalized by the initial half-width, and the results are plotted in figure \ref{fig:runouts}.
Comparison to other work in column collapse is shown in table \ref{tbl:runouts}.
We reiterate that our simulations assume a 3D material in plane-strain, rather than a native  2D material as is used in some of the existing numerical work.
Although the role of the friction angle is debated, we are encouraged that the results from \citet{staron05} (obtained using 2D DEM) are similar, as the friction angle used in those simulations is nearly identical, whereas it is significantly higher in other works.

\begin{figure}
    \centering
    \includegraphics[scale=1.0]{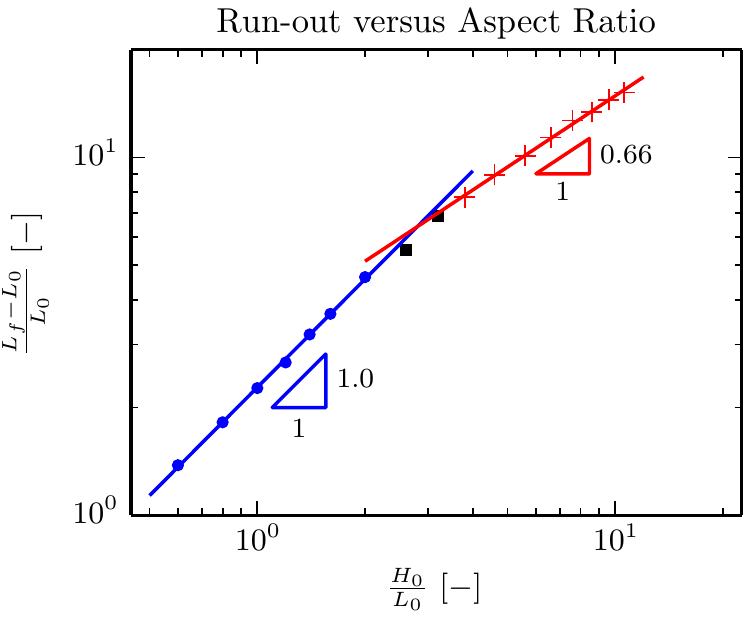}
    \caption{The nondimensional run-out distance as a function of aspect ratio is plotted here.
As is consistent with other literature, two regimes are observed.
When the column is stubby, the run-out distance scales as a linear function of the aspect ratio.
When the column is slender, inertial effects dominate and the run-out distance scales as a sublinear function of the aspect ratio.
The precise form varies significantly, but powers between 0.6 and 0.7 are common. Comparison to other studies, both experimental and numerical, are presented in table \ref{tbl:runouts}.}
    \label{fig:runouts}
\end{figure}

\begin{table}
\centering
\caption{A comparison of numerical and experimental run-out scalings to the current work, similar to that displayed in \citet{mast15} is reproduced here.
We note that our MPM code not only captures the experimentaly observed scaling and transition points, but also matches coefficients well with data from \citet{staron05}, which used a similar friction angle.}
\label{tbl:runouts}
\begin{tabular}{l l | l}
Source & Method & Run-out scaling equation \\
\midrule
\citet{balmforth05} & Experimental &
$\begin{aligned}
\begin{cases} \propto a^{0.9 \pm 0.10} & \text{ if wide channel} \\ \propto a^{0.65 \pm 0.05} & \text{ if narrow slot}
\end{cases}
\end{aligned}$
\\

\citet{staron05} & Numerical (DEM) &
$\begin{aligned}
\begin{cases} 2.5a & \text{ if } a \lessapprox 2.0 \\ 3.25a^{0.71 \pm 0.02} & \text{ if } a \gtrapprox 2.0
\end{cases}
\end{aligned}$
\\

\citet{lube05} & Experimental &
$\begin{aligned}
\begin{cases} 1.2a & \text{ if } a < 1.8 \\ 1.9a^{2/3} & \text{ if } a > 2.8
\end{cases}
\end{aligned}$
\\

\citet{lagree11} & Numerical (Non-Newtonian Model) &
$\begin{aligned}
\begin{cases} 2.2a & \text{ if } a \lessapprox 7.0 \\ 3.9a^{0.7} & \text{ if } a \gtrapprox 7.0
\end{cases}
\end{aligned}$
\\

\citet{mast15} & Numerical (Continuum Plasticity) &
$\begin{aligned}
\begin{cases} (1.01 \pm 0.16)a^{1.52 \pm 0.04} & \text{ if } a \lessapprox 1.5 \\  (1.30 \pm 0.22) a^{0.75 \pm 0.01} & \text{ if } a \gtrapprox 1.5
\end{cases}
\end{aligned}$
\\

Current Study & Numerical (Continuum Plasticity) &
$\begin{aligned}
\begin{cases} 2.28a^{1.00} & \text{ if } a \lessapprox 2.83 \\  3.24a^{0.66} & \text{ if } a \gtrapprox 2.83
\end{cases}
\end{aligned}$
\\
\end{tabular}
\end{table}

}

\subsection{Silo}
Next we conducted simulations of silo discharge. Silos are interesting as they show the ability of the material to disconnect (near the free-fall arch). The Beverloo correlation, found in \citet{beverloo61}, is an empirical rule for relating mass flow rate to the orifice size. The flow rate is independent of the filling height (as long the geometry is large enough compared to the orifice size, see \citet{nedderman92}), which is in stark contrast to Newtonian fluids. The Beverloo correlation in plain strain is given by
\begin{align}
	Q = C\rho\sqrt{g}(D - kd)^{3/2},
\end{align}
where $Q$ is the mass flow rate, $\rho$ is the bulk density, $d$ is the grain diameter, $D = 2r$ is the diameter of the orifice. The parameters $C$ and $k$ are both geometry and material dependent.

\begin{figure}
	\centering
	\begin{tabular}{r H H H}
		Time & $D = 0.06$ m & $D = 0.12$ m & $D = 0.18$ m \\
		\midrule
		0s & \includegraphics[scale=1.0]{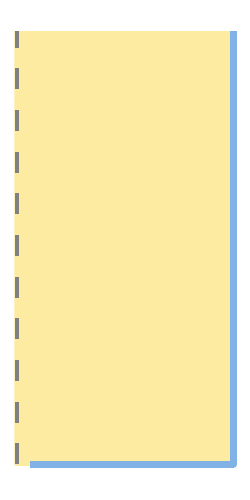} & \includegraphics[scale=1.0]{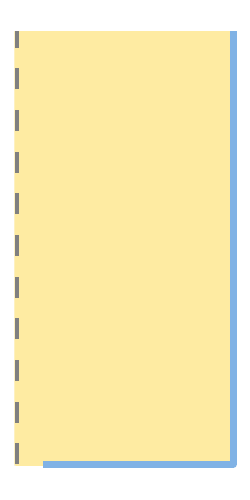} & \includegraphics[scale=1.0]{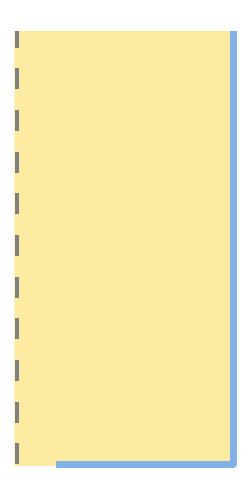} \\
		2s & \includegraphics[scale=1.0]{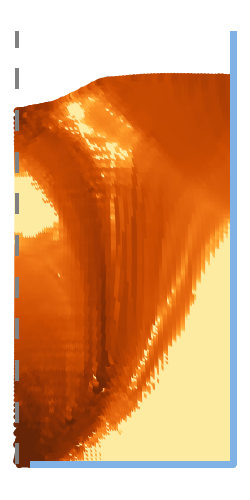} & \includegraphics[scale=1.0]{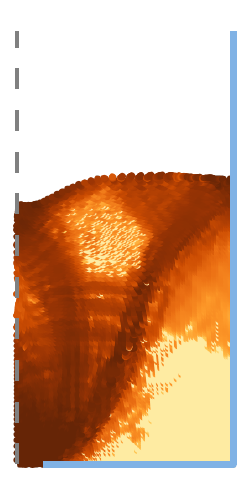} & \includegraphics[scale=1.0]{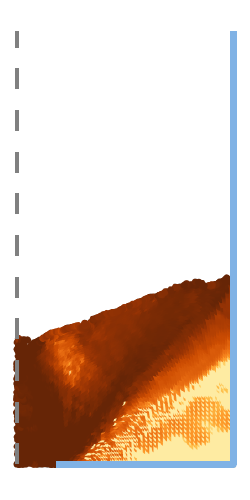} \\
		5s & \includegraphics[scale=1.0]{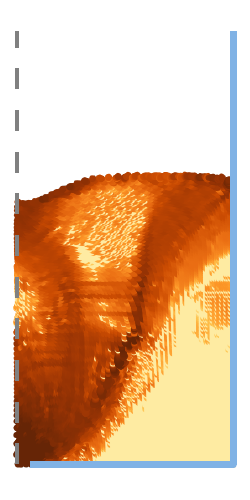} & \includegraphics[scale=1.0]{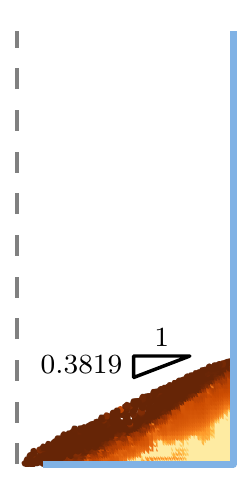} & \includegraphics[scale=1.0]{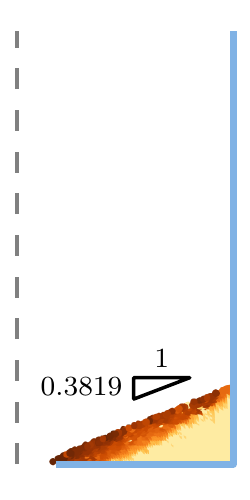} \\
		& \multicolumn{3}{c}{\includegraphics[scale=1.0]{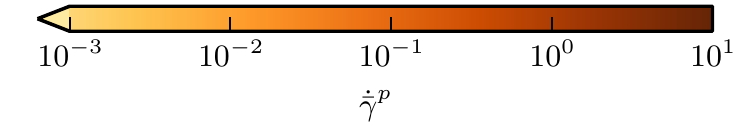}}
	\end{tabular}
	\caption{Parameters for the silo simulations are given in table \ref{tbl:simulation-params}. The behavior of the silos appears physically correct, with the largest orifice size emptying first and in all cases leaving a small amount of material close to the repose angle. The silos initially begin filled to a height of 1.0m, and the base has a width of 1.0m. As before, due to symmetry we only draw half of the geometry.}
	\label{fig:silo-images}
\end{figure}

For our silo simulations, the width of the silo is 1m and the height is 1m (to keep the units consistent, the thickness is also taken to be 1m). The bottom boundary is fully rough, while the sides are frictionless walls, drawn on the right side of each image. The symmetry boundary condition is represented by a dashed line on the left side of each image. We took the orifice size as the distance from the first to last vertically unrestrained node along the bottom left boundary. Although we only simulated half of the silo due to symmetry, we have already taken this into account when presenting the mass and mass flow rate results.

\begin{figure}
\centering
\includegraphics[scale=1.0]{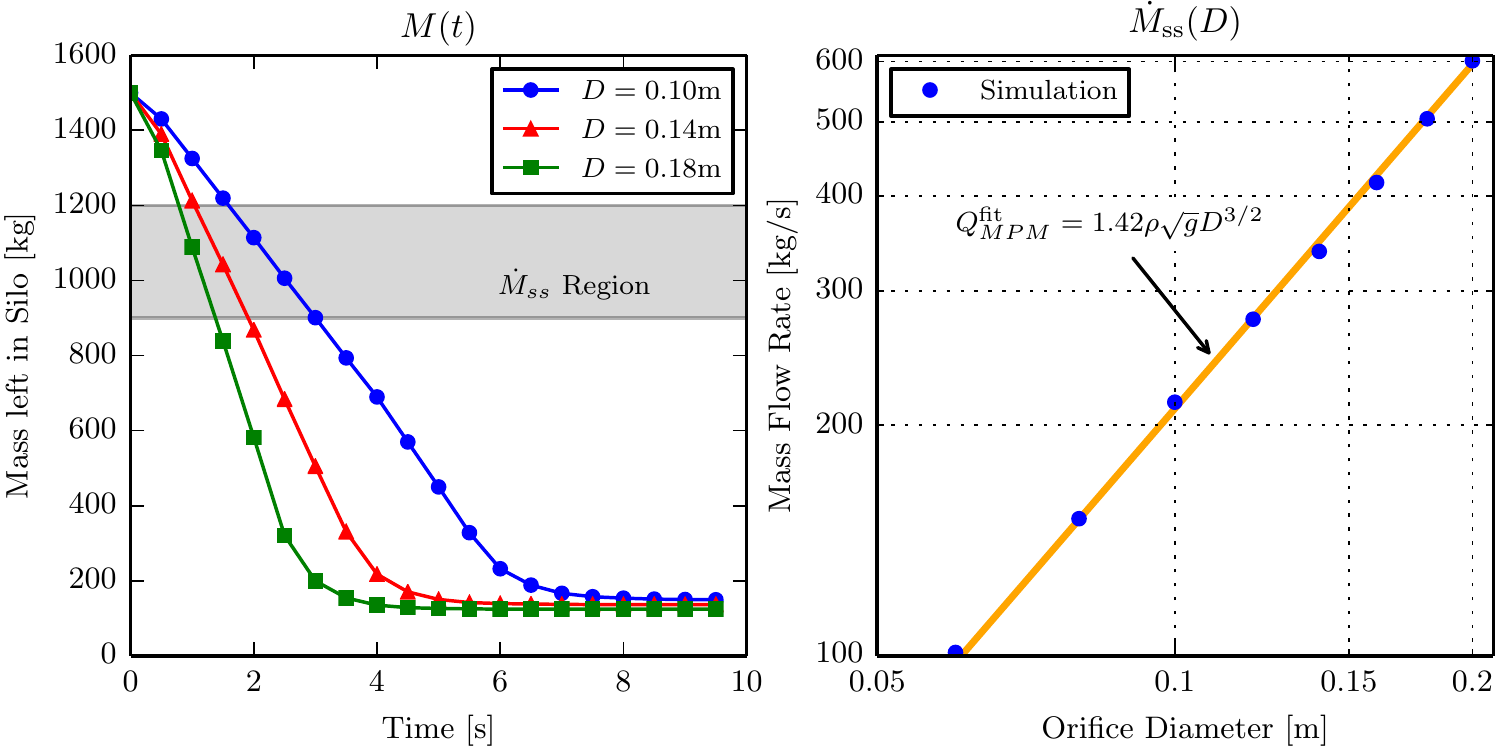}
\caption{Left panel: The mass remaining in the silo is plotted here for a few representative hole sizes. Markers are drawn more sparsely than the actual sampling rate to reduce clutter in the plot. The results indicate a steady flow rate is achieved quickly in the simulation. We label the region where the mass remaining in the silo is between 1200 and 900 kilograms as the steady flow region (to exclude startup effects and to make sure we are in the range of filling heights where Beverloo scaling applies), and we average only data from this region when calculating the flow rate. Right panel: The steady state mass flow rate $\dot{M}_{\mathrm{ss}}$ is plotted here against the hole size in log-log scale. Here the fit $Q^{\mathrm{fit}}_{\mathrm{MPM}}$ is a function of the form $C_{\mathrm{MPM}} \rho \sqrt{g} D^{3/2}$, where $C_{\mathrm{MPM}} = 1.42$. Allowing the power of $r$ to vary as well yields very similar results where the constant is $C_{\mathrm{MPM}} = 1.31$ and the power is $1.47$.}
\label{fig:silo-mass-and-flow-rate}
\end{figure}

We only open the orifice after the gravity reaches the steady value at 0.1s; although in practice we did not see much difference in flow rate compared to opening the orifice at 0s, the stress at material points is better-behaved when allowing the material to settle elastically first.

We plot a few representative samples of silo flow in figure \ref{fig:silo-images}. At 0 seconds, all of the silos are in the same state and are filled to the same height, however at 2 seconds we can already see significant differences in the flow for each orifice size, increasing from 10cm to 14cm to 18cm in diameter from left to right. Again we plot the equivalent plastic shear strain-rate. At 5 seconds, the two largest orifice sizes have nearly finished draining and are becoming static heaps, while the smallest orifice size still has over half of its material remaining.

Additional simulations were carried out to create a set of data for all orifice diameters from 6cm to 20cm in 2cm increments. We looked at each output frame of data, occurring every 1/60 s, and summed the mass of material left in the silo, which we then used to estimate the flow rate during steady flow. A subset of these results is shown in the left panel of figure \ref{fig:silo-mass-and-flow-rate}. We estimated the steady flow region to occur when the mass remaining in the silo was between 1200 and 900 kilograms, and performed a linear regression on the mass as a function of time over this range of data; the slope is then the mass flow rate. The results of these calculations are shown in the right panel of figure \ref{fig:silo-mass-and-flow-rate}, and indeed follow Beverloo scaling. Fitting the data to a function of the form $C_{\mathrm{MPM}} \rho \sqrt{g} D^{3/2}$ yields
\begin{align}
    Q^{\mathrm{fit}}_{\mathrm{MPM}} = 1.42 \rho \sqrt{g} D^{3/2}.
\end{align}
Allowing the power of $r$ to vary as well yields similar results where the constant $C_{\mathrm{MPM}} = 1.31$ and the power on the orifice diameter is $1.47$. More complex constitutive models are needed to determine the parameter $k$, which arises from particle size-effects; the current model uses $k=0$ as expected for a local relation.  It is a major future goal to implement the nonlocal rheology of Kamrin and co-workers \citep{kamrin12, Henann13} within our MPM framework to extract the full Beverloo correlation from a continuum model.

Although similar scaling is observed in other frameworks, such as the incompressible Navier-Stokes solver Gerris in \citet{staron12} and \citet{staron14}, unlike that approach our material model simulates the inertial rheology with a definitive yield criterion and is not restricted to incompressible flow. As extensional disconnection is a small effect during a large portion of the silo flow, however, we find that our flow rates are very similar to those found using Gerris, offering an outside check on our method.

We simulated to a total of 40 seconds for each orifice size. Each simulation took between 2.5 to 8 hours on an Intel i5-2500k using 4 threads. As particles leave the system, less computations must be done in subsequent steps. Due to this effect, faster flowing silos actually take significantly less computational time to simulate. Even accounting for these gains, the large number of particles and the relatively fine grid size require a large amount of time to process; use of an unstructured or refined grid would help decrease the computation time immensely.

\begin{figure}
	\centering
	\begin{tabular}{r C r C}
		0s & \includegraphics[scale=1.0]{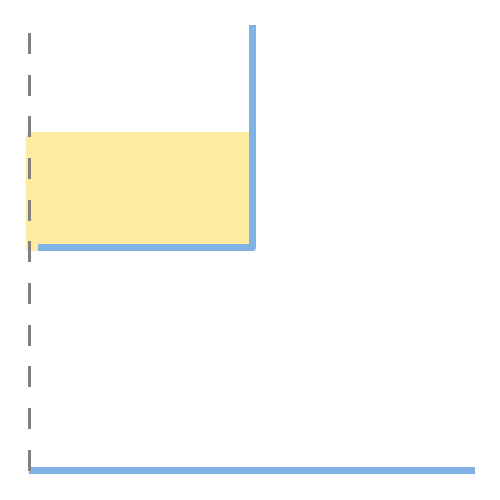} & 1s & \includegraphics[scale=1.0]{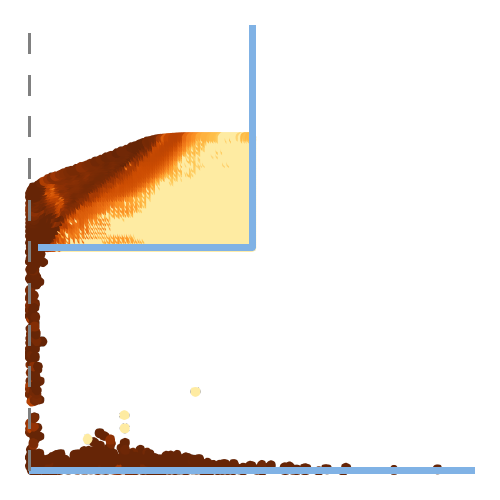} \\
		2s & \includegraphics[scale=1.0]{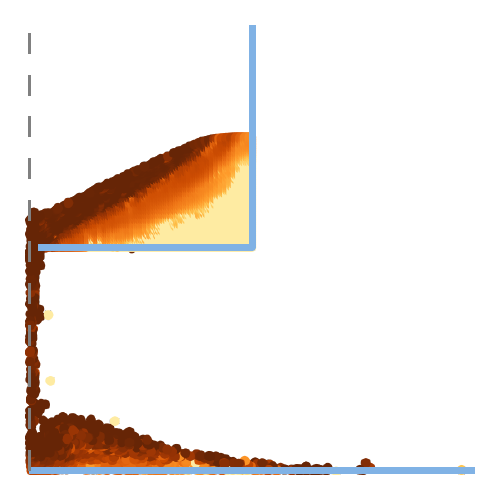} & 5s & \includegraphics[scale=1.0]{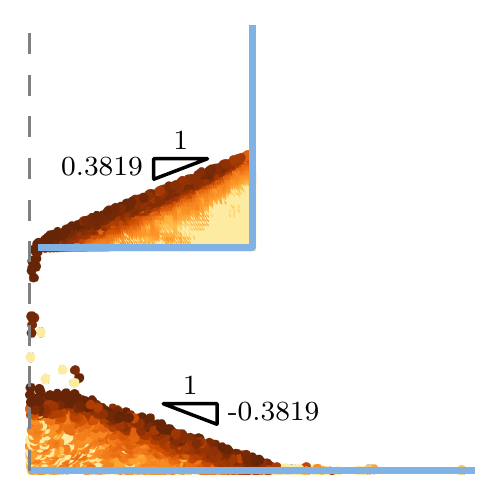} \\
	\end{tabular}
	\begin{tabular}{H H H H}
		\midrule
		\includegraphics[scale=1.0]{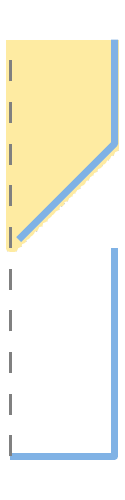} &
		\includegraphics[scale=1.0]{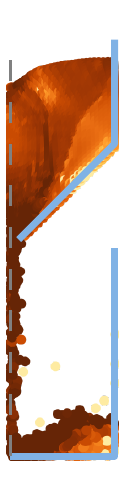} &
		\includegraphics[scale=1.0]{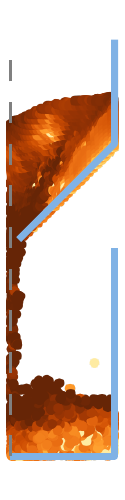} &
		\includegraphics[scale=1.0]{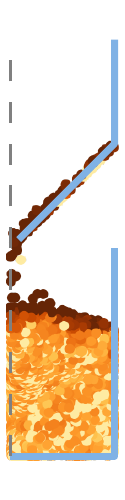} \\
		0s & 1s & 2s & 5s
	\end{tabular}
	\includegraphics[scale=1.0]{figs/standalone_colorbar}
	\caption{Parameters for the hourglass simulations are given in table \ref{tbl:simulation-params}. The dashed boundary indicates the symmetry condition. Material points are free to slide along walls without friction, except in the case of the bottom surface, which is fully rough. In both cases the radius of the orifice's image on the x-axis is 0.02m.}
	\label{fig:hourglass-images}
\end{figure}

\subsection{Drainage and Reconsolidation}
The hourglass-like examples in this section highlight the capability of the scheme to model media that flows plastically, undergoes granular phase change to a disconnected state, reconsolidates, and supports stress again as an elasto-plastic body. Other continuum solvers would have difficulty simulating the entirety of this process, but this is tractable in MPM. They are performed in the same manner as the silo simulations; only the boundary and initial configuration is changed.

\begin{figure}
	\centering
	\begin{tabular}{r C r C}
		0s & \includegraphics[scale=1.0]{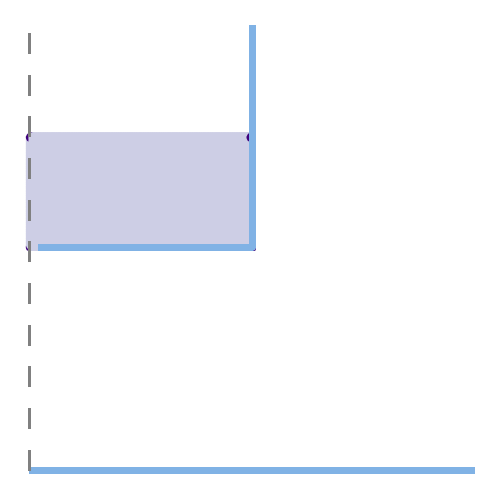} & 1s & \includegraphics[scale=1.0]{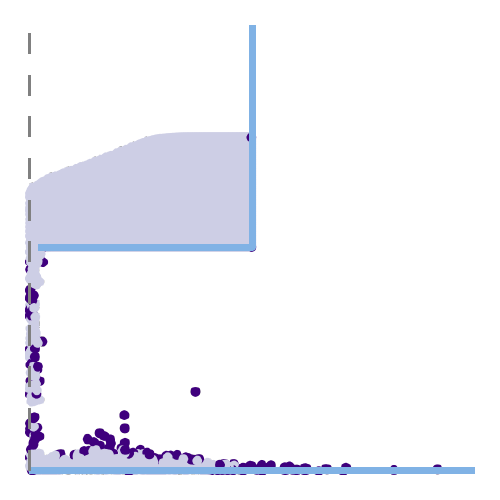} \\
		2s & \includegraphics[scale=1.0]{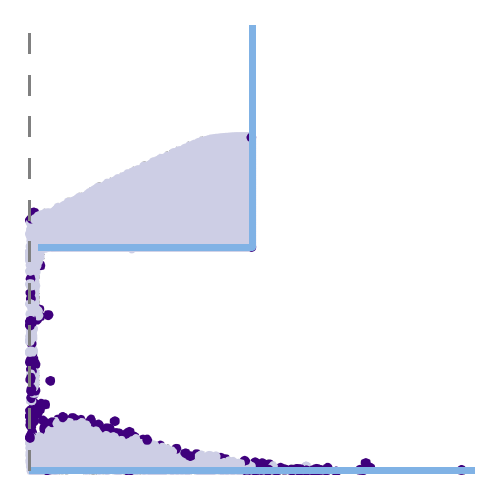} & 5s & \includegraphics[scale=1.0]{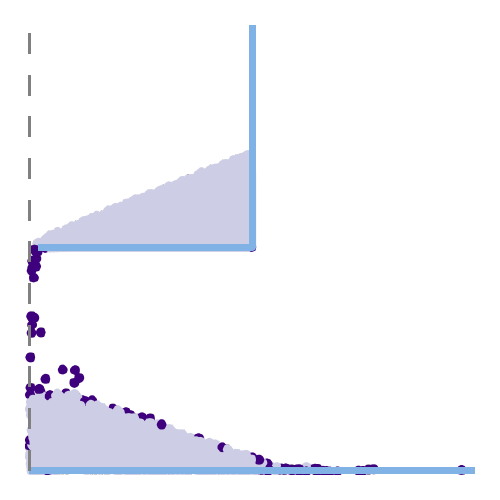} \\
	\end{tabular}
	\begin{tabular}{H H H H}
		\midrule
		\includegraphics[scale=1.0]{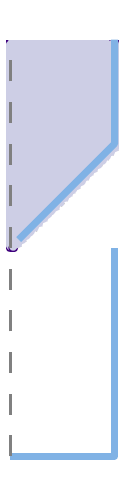} &
		\includegraphics[scale=1.0]{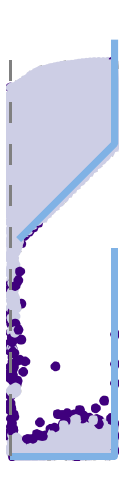} &
		\includegraphics[scale=1.0]{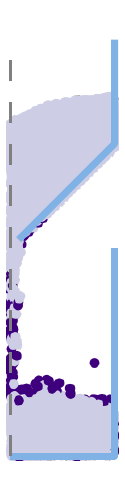} &
		\includegraphics[scale=1.0]{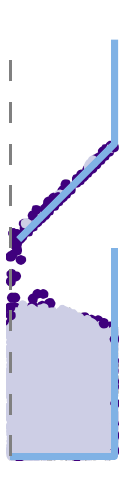} \\
		0s & 1s & 2s & 5s
	\end{tabular}
	\includegraphics[scale=1.0]{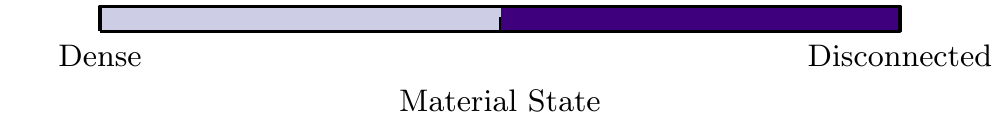}
	\caption{The same simulations as in figure \ref{fig:hourglass-images} are presented here, however in these images we plot the state of the material throughout the simulation. As expected, material points which are splashing or in the column of falling particles are in the disconnected state, while material points which support stress are in the dense state.}
	\label{fig:hourglass-state-images}
\end{figure}

For both examples, the height from the base to the orifice is 0.5m. The radius of the orifice's image on the x-axis in both cases is 0.02m. Since we are using a Cartesian grid, the sloped boundary condition is implemented by fixing a staircase pattern of nodes and applying frictionless conditions.

The initial width of the flat hourglass is 1.0m, while the width of the sloped hourglass is 0.5m. The fill height of the top hourglass is 0.25m, and the fill height of the sloped hourglass is 0.25m. The bottom boundary condition is a fully rough wall in both cases; a frictionless wall is present at a radius of 0.25m in the sloped hourglass while no side wall is used in the flat hourglass.

The hourglasses were simulated to a total of 10 seconds each, although both cases are nearly at the final configuration by the 5 second mark. Figure \ref{fig:hourglass-images} shows the results of each simulation. The top half of the table shows the hourglass with a flat divider at 0 seconds (top left), 1 second (top right), 2 seconds (bottom left), and 5 seconds (bottom right). In the bottom half of the table, we see the results from the sloped hourglass at the same times from left to right. The equivalent plastic shear strain-rate is plotted in both sets of images. As with the heaps, we see that the simulations settle on a repose angle less than or equal to $\arctan (\mu_s)$.

We can look at the state of the particles during the simulation to confirm that our algorithm is using the appropriate behavior in the different regimes. In order to plot the state, we first calculate the density on the nodal level. To do this, we project the mass to the nodes as usual. However, we also assign a nodal volume, which is defined as the fraction of filled neighboring elements times the volume of the Voronoi cell around the node (with a regular Cartesian grid, this volume is the same as the volume of each element, excepting those nodes on edges which have half the elemental volume and corners which have a quarter of the elemental volume). The nodal density is then defined as the nodal mass over the nodal volume. We then project this density back to the material points using the shape functions. This is identical to the method presented in \citet{wieckowski04}, except we only do this in post-processing (the density during the simulation is calculated on the material points using the exponential map). Plotting this nodal-based density instead of the material point density allows us to see the average collective behavior of the elements, to represent what behavior enters the stress-divergence of the momentum update. The values below $\rho_c$ are in the disconnected state, while above $\rho_c$ the material acts as an elasto-plastic body (i.e. the `dense' state).

Figure \ref{fig:hourglass-state-images} shows the instantaneous state using the nodal-density plotting technique. We see that the behavior of the particles follows our intuition, in that material points which are splashing or in free-fall are in the disconnected state, while material points which support load are in the dense state.

The hourglass simulations took approximately 50 minutes each on an Intel i5-2500k using 4 threads. Although the geometries are different, the number of material points is largely comparable between the simulations and the run time does not differ significantly.

\subsection{Inclined Plane}
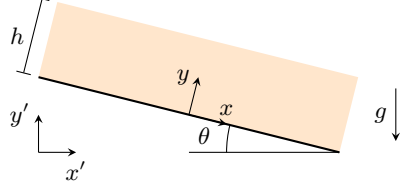
\begin{figure}
\centering
\begin{tikzpicture}
\path[draw=black, thin, ->, >=stealth] (0cm, 0cm) -- (0.5cm, 0cm) node[pos=1,below] {$x'$};
\path[draw=black, thin, ->, >=stealth] (0cm, 0cm) -- (0cm, 0.5cm) node[pos=1,left] {$y'$};

\path[draw=black, thin, ->, >=stealth] (4.75cm, 0.85cm) -- (4.75cm, 0.15cm) node[pos=0.5,left] {$g$};

\path[fill,color=orange!20] (0cm, 1cm) -- (4cm, 0cm) -- (4.25cm, 1cm) -- (0.25cm, 2cm);
\path[draw=black, thick] (0cm, 1cm) -- (4cm, 0cm);

\path[draw=black, thin, ->, >=stealth] (2cm, 0.5cm) -- (2.5cm, 0.375cm) node[pos=1,above] {$x$};
\path[draw=black, thin, ->, >=stealth] (2cm, 0.5cm) -- (2.125cm, 1.0cm) node[pos=1,left] {$y$};

\path[draw=black, thin, |-|] (-0.2cm, 1.05cm) -- (0.05cm, 2.05cm) node[pos=0.5,left] {$h$};

\path[draw=black, thin] (4cm, 0cm) -- (2cm, 0cm);
\path[draw=black, thin, -] (2.5cm, 0cm) arc(180:166:1.5cm);
\node[above left] at (2.4cm, 0cm) {$\theta$};
\end{tikzpicture}
\caption{A schematic of the setup for the inclined plane problems. The global axes are $x'$ and $y'$, while the axes $x$ and $y$ are parallel and perpendicular to the inclined plane respectively. The bottom of the inclined plane is a fully rough boundary. }
\label{fig:inclined-plane}
\end{figure}

\begin{figure}
	\centering
	\begin{tabular}{C C C}
		\includegraphics[scale=1.0]{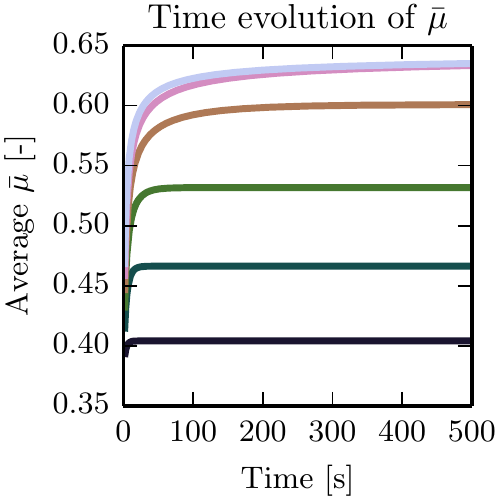} & 
		\includegraphics[scale=1.0]{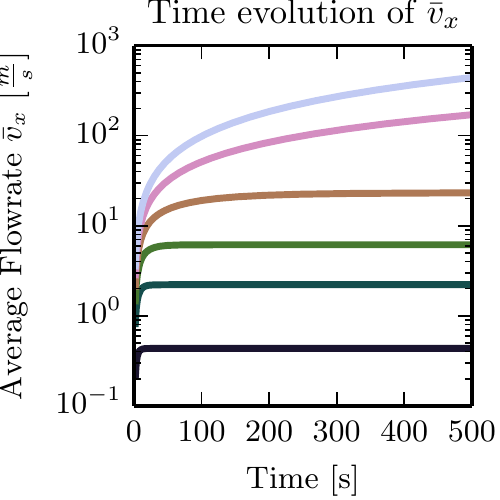} & 
		\includegraphics[scale=1.0]{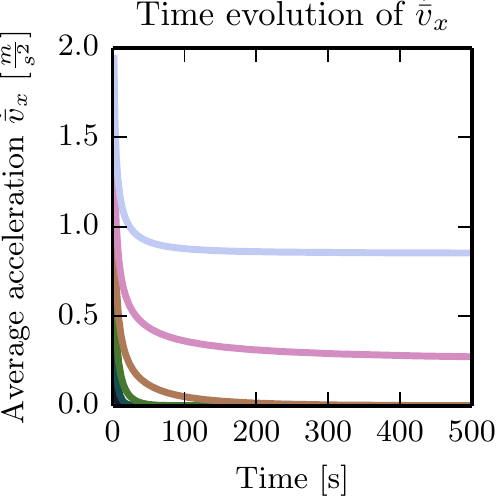}
	\end{tabular}
	\includegraphics[scale=1.0]{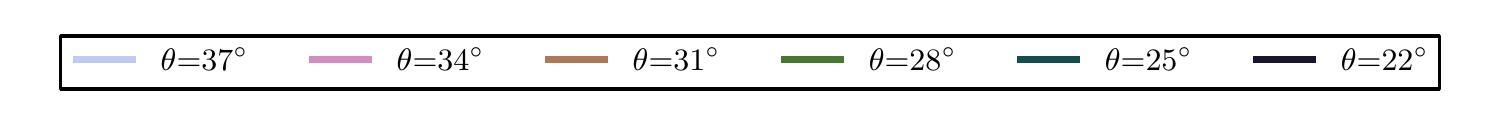}
	\caption{ Left panel:  Evolution of the stress ratio. For tilt angles between $\arctan (\mu_s)$ and $\arctan (\mu_2)$, the friction coefficient evolves to match the value predicted by quasistatic analysis, $\mu = \tan \theta$. For tilt angles above $\arctan (\mu_2)$, $\mu$ approaches $\mu_2$ per the restrictions of the constitutive relation. Center panel:  Depth-averaged speed for multiple tilt angles. For tilt angles between $\arctan (\mu_s)$ and $\arctan (\mu_2)$, the speed asymptotes to a finite value. When the tilt angle is above $\mu_2$, the value of $\mu$ the flow accelerates indefinitely. Right panel: Depth-averaged acceleration. As expected, the acceleration drops to zero for angles between $\arctan (\mu_s)$ and $\arctan (\mu_2)$, while above $\arctan (\mu_2)$ the acceleration asymptotes to a nonzero value.}
	\label{fig:incline-accel}
\end{figure}
The inclined plane flow examples serve to highlight the rate-dependent nature of the dense flow model. Using a rate-independent constitutive equation such as the standard Drucker-Prager yield criterion will not capture a unique steady flow profile over a non-zero range of incline angles.  The inertial rheology, which we use here, is able to capture this important phenomenon, which arises in many industrial contexts.  In fact, the observation that granular flows down a rough incline achieve a steady-state velocity for inclinations  $\theta$ between $\theta_1=\arctan(\mu_s)$ and $\theta_2=\arctan(\mu_2)$ was a major motivator in the historical development of the inertial rheology.

The problem is set up as a periodic inclined plane placed at an angle $\theta$ with respect to the horizontal (see figure \ref{fig:inclined-plane} for geometric inputs). The material on top of the plane begins in an initially stress free state and over time develops a unique steady state profile.

In our simulations, the height of the material perpendicular to the inclined plane is 20cm. As in the other simulations, gravity is not applied instantaneously but ramps up to its final value over the first tenth of a second. Because the solution is invariant in $x$, we simulate a single column of elements, using periodic boundary conditions on the left and right sides of the element.  The simulations were run at angles of 19 degrees to 34 degrees from the horizontal in 1 degree increments. The analytical solution for the steady-state velocity profile, obtained by integrating through the chosen rheology, yields the Bagnold profile given by
\begin{align}
	\frac{v_x(y)}{\sqrt{gh}} = \frac{2}{3} \xi \frac{\tan \theta - \mu_s}{\mu_2 - \tan \theta} \sqrt{\rho_s \Phi \cos(\theta)} \frac{h^{3/2} - (h - y)^{3/2}}{h^{1/2}}.
	\label{eqn:bagnold-profile}
\end{align}
The Bagnold profile has been observed rather clearly in DEM simulations of inclined flows \citep{silbert01, silbert03} \revised{when the height of the layer is sufficiently large compared to the grain size}.
The material parameters above are identical to those we have been using in the other simulations.
Integrating equation \eqref{eqn:bagnold-profile} over the entire height and dividing by $h$ yields the depth-averaged velocity, given by
\begin{align}
\frac{\bar{v}_x}{\sqrt{gh}} = \frac{2}{5} \xi h \frac{\tan \theta - \mu_s}{\mu_2 - \tan \theta} \sqrt{\rho_s \Phi \cos(\theta)}
	\label{eqn:depth-averaged-velocity}.
\end{align}
These equations only apply for angles between $\arctan (\mu_s)$ and $\arctan (\mu_2)$. If the angle supplied is lower than $\arctan (\mu_s)$, there should be no flow. At angles above $\arctan (\mu_2)$, the system cannot find a steady state as the constitutive model renders it impossible to achieve equilibrium, and the material will continually accelerate.
\revised{The results indicate very high velocities at high tilt angles due to the thickness of the layer, however they match the analytical predictions of the $\mu(I)$ model (for an infinitely long chute with no drag sources).}
\begin{figure}
	\centering
	\begin{tabular}{J J} \includegraphics[scale=1.0]{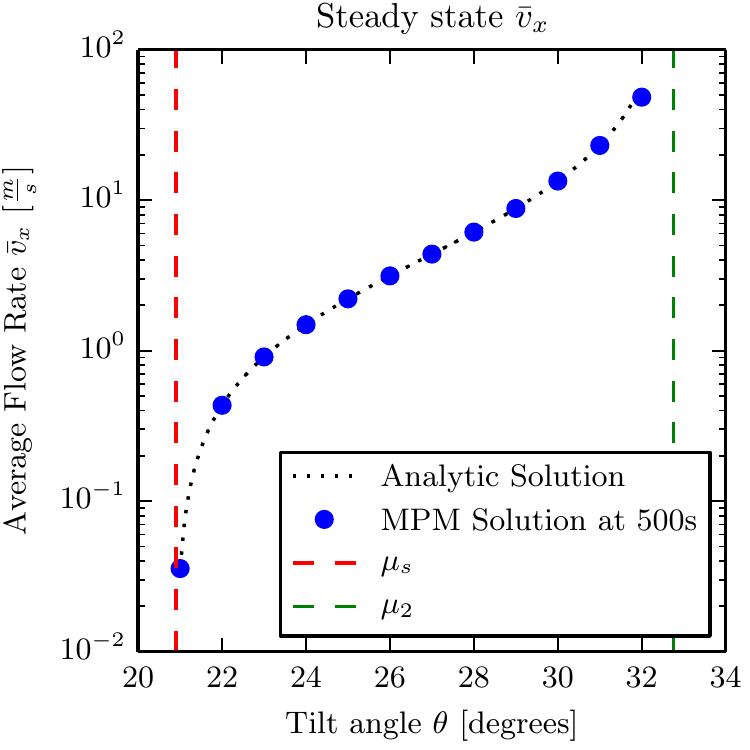} & \includegraphics[scale=1.0]{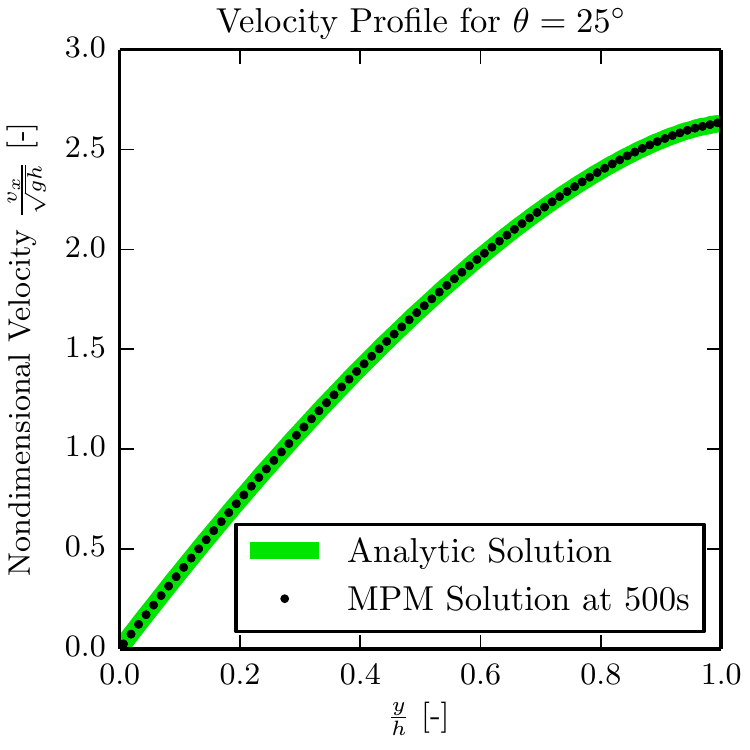}
	\end{tabular}
	\caption{Left panel: The depth-averaged flow rate for multiple tilt angles. Right panel: The velocity profile from the MPM simulation for $\theta = 25^\circ$ in comparison to the analytical solution.}
	\label{fig:incline-flowrate}
\end{figure}
We first confirm that the system cannot find a steady state velocity profile for angles above $\arctan (\mu_2)$. Figure \ref{fig:incline-accel} shows the evolution of the depth-averaged $\mu$ in the leftmost figure. For angles above $\arctan (\mu_2)$ the value of $\mu$ approaches $\mu_2$, but is prevented from exceeding it through the constitutive model. This necessitates that the material cannot satisfy quasi-static balance and must continuously accelerate. The middle and right images in figure \ref{fig:incline-accel} show the time evolution of average velocity and average acceleration respectively, and also confirm that a steady state does not exist for angles above $\arctan (\mu_2)$ (in our case, this is approximately $32.8$ degrees).

We also confirm that our MPM simulations are able to reproduce the analytical solutions for both the Bagnold profile and the depth-averaged flow rate as a function of angle in figure \ref{fig:incline-flowrate}. Although not shown, as expected we also found that angles less than $\arctan (\mu_s)$ -- approximately $20.9$ degrees in our case -- did not flow (to within machine precision) after the initial loading.

Although this is a special case, note that the material points in the simulations with larger tilt angles reach shear strains with magnitudes exceeding $10^5$. Attempting a similar analysis with a naive finite element implementation would not be possible.

In order to ensure the ability to reach steady-state, especially for tilt angles close to $\arctan (\mu_2)$, we simulated a total of 500s for each simulation.  Each simulation took approximately 5 hours on an Intel E5645 using one thread (increasing threads does not help significantly as the amount of work per step is small in this case).

\begin{figure}
	\centering
	\begin{tabular}{H H H H H}
		\multicolumn{2}{C}{\includegraphics{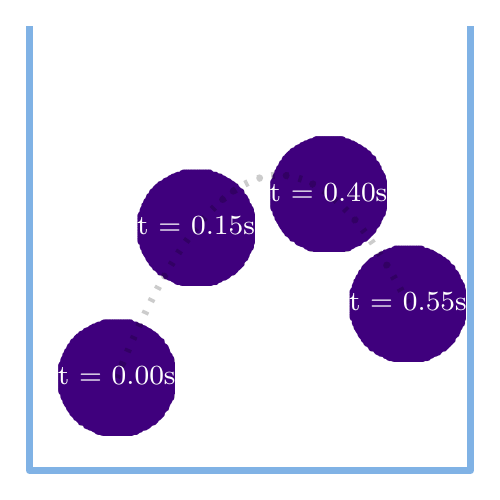}} &
		\includegraphics{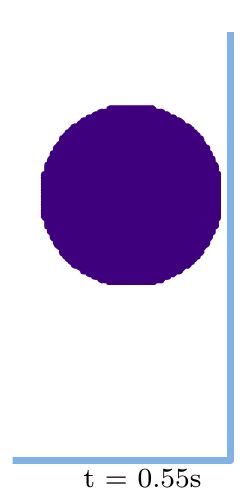} &
		\includegraphics{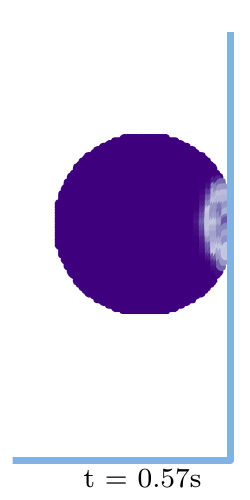} &
		\includegraphics{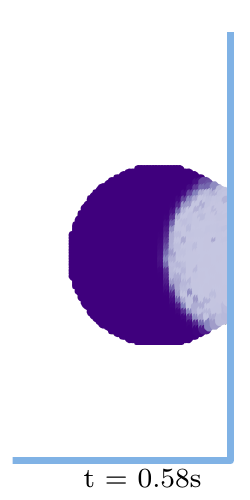} \\
		\includegraphics{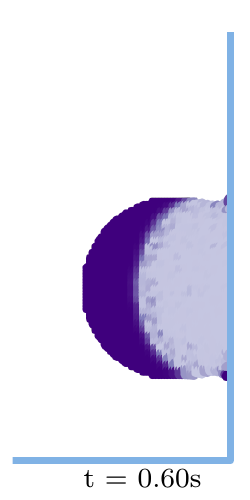} &
		\includegraphics{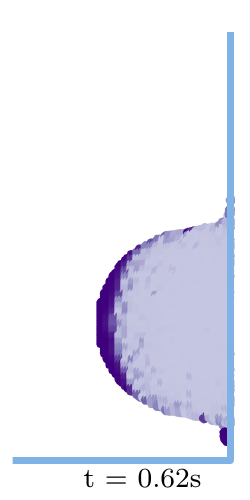} &
		\includegraphics{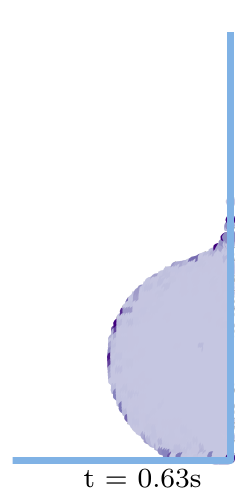} &
		\multicolumn{2}{C}{\includegraphics{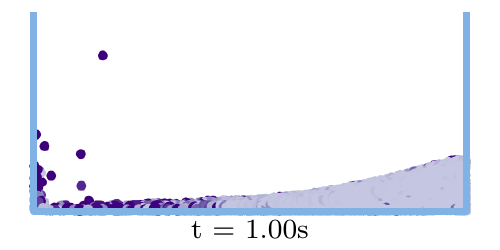} \includegraphics{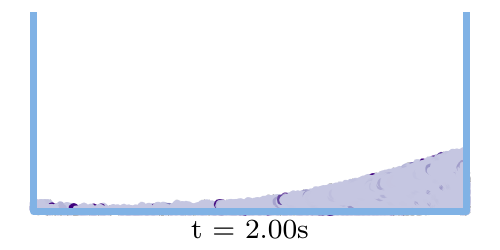}}
	\end{tabular}
	\includegraphics{figs/standalone_state_colorbar}
	\caption{Granular slug simulation:  The projectile initially follows the parabolic trajectory given by classical kinematics (top left). When it begins to interact with the wall, the material undergoes densification, which is captured through the series of images (zoomed-in) along the remainder of the top row and the left side of the bottom row. Finally, the material comes to rest and is able to support stress with a density higher than the critical value, forming a static heap (bottom right).} 
	\label{fig:projectile}
\end{figure}

\subsection{Granular Slug}

To further demonstrate our method's ability to capture phase changes, we model a slug of loose granular material that is thrown against a rigid wall, causing it to reconsolidate and then plastically collapse into a heap on the floor.  The slug begins as a disc of disconnected ($\rho < \rho_c$) material and is given an initial velocity. The components of the initial velocity are 1.2 m/s in the positive x direction and 3.0 m/s in the positive y direction. For this simulation alone we did not ramp gravity, and instead it is held at the fixed value of 9.8 m/$\mathrm{s}^2$ throughout the entire duration.

Figure \ref{fig:projectile} shows the results of the projectile simulation. We see that the projectile initially follows a parabolic path as predicted from classical kinematics. However, as the material begins in a disconnected state, when the projectile impacts the rigid boundary it undergoes densification. This transition is captured in the propagating front shown in the figure. Finally, as the material collapses into a heap, we see that it remains in the dense state as it must now support stress.  As before, the walls are modeled with perfect slip conditions and the floor is modeled fully-rough.  The projectile simulation further highlights the ability of our MPM scheme to represent large inhomogeneous deformations  of granular media while maintaining realistic results.

\section{Conclusion}
We have developed a model and continuum simulation procedure for granular materials, specialized to handle the solid-like (elastic), liquid-like (plastic), and gas-like (disconnected) behaviors that these materials frequently switch between. By tracking the material density, it allows the use of an elasto-plastic behavior when the material is dense while being able to transition into a disconnected state when the material is separated and should not support stress, and vice versa. Our stress update algorithm, which is designed to handle these phase transitions, has been implemented in the context of an MPM framework, which naturally allows for large inhomogeneous deformations, and the combination is able to successfully replicate a variety of extremely large-strain phenomena (e.g. silo drainage, long-time inclined chute flows) which would not be possible in a simple FEM implementation. The physicality has been validated a number of ways: through the approach to a static repose angle in collapsing heaps, the Beverloo scaling of silo drainage, and the proper range of steady flow inclinations (and corresponding flow profiles) in inclined chute flows.  We have also demonstrated the ability of the scheme to handle transitions into and out of the dense phase.

A natural extension would be the verification and validation of the method in fully 3D cases. Another useful component would be the implementation of moving Lagrangian intruders within a flow (e.g. for mixing problems). Although a solution which uses `boundary material points' is provided in \citet{Ma2009}, this effectively blurs the boundary condition over an element size. On the constitutive side, in the dense phase we can append a model for the transient plastic flow effects, such as shear strengthening/weakening and shear-dilation, per a critical-state soil mechanics model \citep{schofield68}. Finally, a major goal is the implementation of nonlocal granular models (e.g. that of \citet{kamrin12}) in a similar manner, which would be an important component of a predictive continuum tool for simulating general granular flows. 

\section{Acknowledgement}
KK would like to acknowledge support from NSF-CBET-1253228 and the MIT Department of Mechanical Engineering.

%
%
\bibliographystyle{jfm}
\bibliography{./mpm_references,./scipy,./added_references}

\appendix
\section{Derivation of Equation of State and Relation to Kinetic Theory}
\label{sec:kinetic-theory}
\revised{
To examine our expression for the equation of state, we additively decompose the inverse density into elastic and flow contributions,
\begin{align}
    \frac{1}{\rho} = \frac{v_e}{m} + \frac{v_f}{m}.
\end{align}
The elastic component of volume relates to how much the grains are compacted in a densely packed body, which is given by a $\tr \mathbf{E}$ term.
The flow component of volume relates to the bulk volume change due to grain structure physically dilating as grains move around each other.
In reality, this includes effects such as Reynolds dilation, but for our approximation we will take the flowing volume as a Heaviside function, which is unity when $\rho < \rho_c$ and zero otherwise.  The accuracy of this approximation will be analyzed in a moment.

For computing the elastic volume, we simply take the pressure in tension to be zero (since grains which are not in contact cannot transmit stresses).
In compression, the slope of the pressure to elastic volume change curve is proportional to the bulk modulus.

The sum of these two volumes can now be used to write an equation of state for the pressure as a function of the bulk density.
However, the functional form necessarily has different domains due to this decoupling assumption, which creates a kink in the pressure as shown in equation \eqref{eqn:eos}. Taking the time derivative of the equation of state yields an expression we may use to update the stress, which we will expand upon shortly.  We can differentiate the equation of state to obtain the bulk modulus, given by
\begin{align}
K = \rho \frac{\partial p}{\partial \rho} = \begin{cases}
        0 & \text{ if } \rho < \rho_c \\
        \frac{K_c \rho_c}{\rho}  & \text{ if } \rho \ge \rho_c \end{cases}.
\end{align}
Thus the parameter $K_c$ is seen to be the limit of the bulk modulus of the material at $\rho_c$ when approached from above. The time derivative of equation \ref{eqn:eos} is given by
\begin{align}
    \dot{p} = \begin{cases}
        0 & \text{ if } \rho < \rho_c \\
        K_c \frac{\rho_c \dot{\rho}}{\rho^2} & \text{ if } \rho \ge \rho_c \end{cases},
\end{align}
and we can replace the $-\dot{\rho} / \rho$ term with $\tr \velgrad$ due to the local form of mass balance in equation \eqref{eqn:local-density}.
Similarly, we can rearrange equation \eqref{eqn:eos} to obtain the term $K_c \rho_c / \rho = K_c - p$.
Substitution of these quantities into the time derivative yields
\begin{align}
    \dot{p} = \begin{cases}
        0 & \text{ if } \rho < \rho_c \\
        -K_c \tr \velgrad + p \tr \velgrad & \text{ if } \rho \ge \rho_c \end{cases}.
\end{align}
In the linear elastic limit the pressure is much smaller than the bulk modulus, so we can neglect this term and take the time derivative as $\dot{p} = -K_c \tr \velgrad$, which is the same as the expression obtained via linear elasticity with constant bulk modulus equal to $K_c$.
This rule is natively incorporated within the objective stress rate and density conditional of Eqs \ref{eqn:constitutive-cases} and \ref{eqn:constitutive-stress-free}.

We are aware that the form of the flowing volume is approximate; we will now address when the approximation is valid and how to determine the level of errors incurred.
Since we take the pressure as zero below $\rho_c$, the approximation is valid when the true pressure is small compared to a characteristic pressure in the system for values of $\rho$ less than $\rho_c$.
In the general case, there are two regimes of flow which we can consider as model cases.
The first case is steady-state shear flow, and the other is volumetric compression or expansion.
First note that in the case of a steady-state shear flow, we can use the $\Phi(I)$ relation to determine the pressure given a shear strain rate and density.
For the volumetric compression and expansion, we can use kinetic theory to model the granular gas and derive an expression for the pressure given a volumetric strain rate and density.

In both cases if the pressure calculated at $\Phi_c = \rho_c / \rho_s$ is small compared to a reference pressure in the system, the approximation is valid.

\begin{figure}
\centering
\begin{tabular}{c c}
    \includegraphics[scale=1.0]{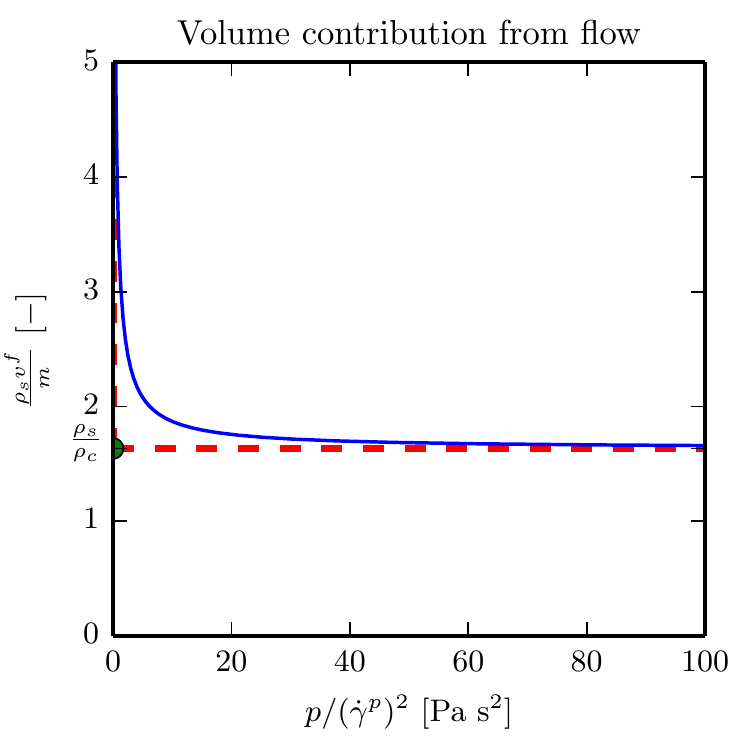} &
    \includegraphics[scale=1.0]{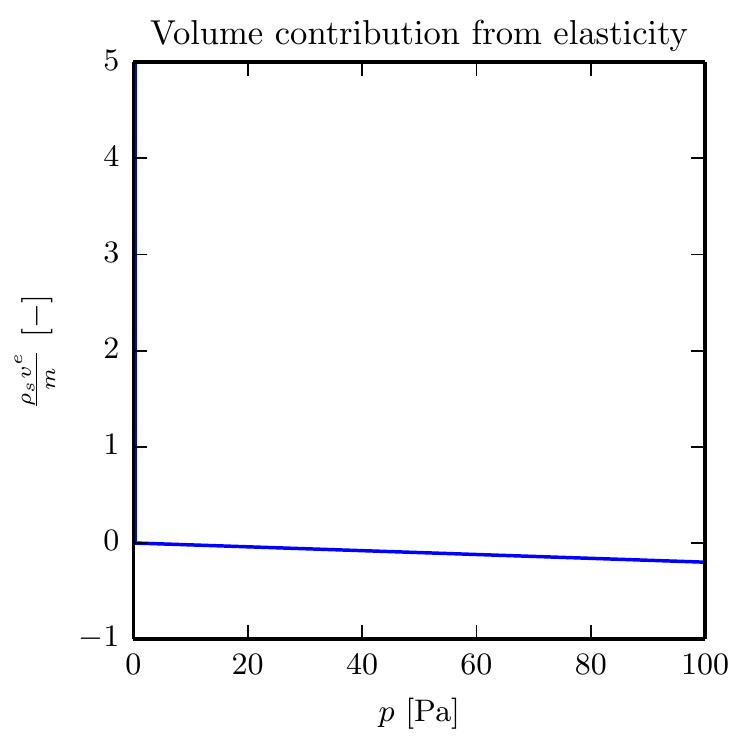} \\
\end{tabular}
\caption{
The different volume contributions are shown here.
We approximate the actual $\phi(I)$ function, shown in the left panel by a solid line, by a much sharper step function, which can take on any value above a critical inverse packing fraction at zero pressure.
At the critical value, the pressure may take on any positive value.
This approximate function is then used as the flowing volume.
The elastic volume is simply given by linear elasticity in compression and zero in tension; the slope of the solid line in the left panel is the negative inverse of the bulk modulus of the material.
The total inverse density is the sum of these two responses scaled by the solid density of the grains.}
\label{fig:inverse-packing}
\end{figure}

To examine the validity of the stress-free approximation in the gas phase (when $\rho < \rho_c$), we reproduce the arguments and notation of \citet{jenkins12} here.
The equation of state for the pressure is given by
\begin{align}
    p = 4 \rho \tilde{G} \tilde{F} T.
\end{align}
Here, $\tilde{G} \equiv \Phi (1 - \Phi / 2) / (1 - \Phi)^3$ when $\Phi < 0.49$ and $\tilde{G} \equiv 0.63 \Phi / (0.6 - \Phi)$ when $\Phi \geq 0.49$,
$\tilde{F} = (1 + \epsilon)/2 + 1/(4\tilde{G})$ ($\epsilon$ is the effective coefficient of restitution), and $T$ is the granular temperature (as before, $\rho$ is the density and $p$ is the pressure).
This can be further decomponsed into normal and tangenetial restitution coefficients, but for our purposes it suffices to consider $\epsilon$ simply as a value between zero and one.
The pressure grows quickly when approaching $\Phi = 0.60$ due the division in the expression for $\tilde{G}$.
Above this value, the kinetic stresses are dominated by enduring elastic contacts between grains.
Below this value, the expression for pressure is linearly dependent on the granular temperature $T$; thus, as long as the temperature remains small, we can approximate the pressure as zero in comparison to the stresses experienced when the material is packed together.

We can take the expression for pressure and immediately derive an expression for the $\Phi(I)$ relation.
Using the argument from \citet{dacruz05}, the temperature is related to the fluctuations in velocity.
In shear the temperature is given by
\begin{align}
T_s &= \dot{\gamma}^2 d^2 \frac{1}{9I}.
\end{align}

The shear contribution to the flowing volume is then
\begin{align}
\frac{v_f}{m} &= \frac{4 \lambda \tilde{G}(\Phi) + 1 }{p} \left( \dot{\gamma}^2 d^2 \frac{1}{9I} \right)
\end{align}
where $\lambda = (1 + \epsilon) / 2$.

First we use $\rho = \rho_s \Phi$ and collect terms (recalling that $I = \dot{\gamma} \sqrt{\rho_s d^2} / \sqrt{p}$) to arrive at
\begin{align}
(0.6 - \Phi) \frac{9}{I} = \left( 2.52 \lambda - 1\right) \Phi^2 + 0.6 \Phi
\end{align}

Since the inverse density is related to the inverse packing fraction, we solve this quadratic equation for inverse packing fraction $1/\Phi$ to arrive at
\begin{align}
\frac{1}{\Phi} = \frac{1}{2} \left( \frac{I}{9} + \frac{1}{0.6} + \sqrt{ \left( \frac{I}{9} + \frac{1}{0.6} \right)^2  + \frac{4I}{5.4} (2.52 \lambda - 1) } \right).
\end{align}

Given a shear strain rate, we can then compute the pressure as a function of packing fraction to be
\begin{align}
p(\Phi) = \frac{\rho_s \Phi^2}{81} \left( \frac{2.52 \lambda \Phi}{0.6 - \Phi} + 1 \right)^2 \dot{\gamma}^2 d^2
\end{align}
if this was the only contribution to the volume.
When we add the elastic part, we must solve the coupled system of equations
\begin{align}
\frac{1}{\Phi} &= \frac{1}{\Phi_e} + \frac{1}{\Phi_f} \\ 
-K_c \Phi_e &= \frac{\rho_s \Phi_f^2}{81} \left( \frac{2.52 \lambda \Phi_f}{0.6 - \Phi_f} + 1 \right)^2 \dot{\gamma}^2 d^2,
\end{align}
where the physical solution occurs when $\Phi_e < 0$ and $\Phi_f > 0$.
The pressure at this packing fraction may then be calculated via the elastic packing fraction by the relation $p = -K_c \Phi_e$.
If we wish to consider a small range of density variation (given by a small range in packing fraction $\delta$) as constant density, we can take $\Phi + \delta$ instead of $\Phi$ in the first equation.

To obtain the contribution to flowing volume from volumetric strain rates, we use the temperature evolution equation from \citet{jenkins12}.
This is used to obtain a maximum temperature as a function of volumetric strain rate
The balance of fluctuation energy yields a differential equation for the temperature given by
\begin{align}
    (3/2)\rho \dot{T} = \tr{(\cauchystress \symvelgrad)} - \sDiv \mathbf{q} - \Gamma,
\end{align}
where $\mathbf{q}$ is the flux of the fluctuational energy and $\Gamma$ is the collisional dissipation.
We consider here the case that there is no fluctuation energy flux entering the system, so the equation reduces to
\begin{align}
    (3/2)\rho \dot{T} = \tr{(\cauchystress \symvelgrad)} - \Gamma.
    \label{eqn:kinetic-no-q}
\end{align}
The form of $\Gamma$ is given by
\begin{align}
    \Gamma = \frac{12}{\sqrt{\pi}} \frac{\rho \tilde{G}}{d} (1 - \epsilon^2) T^{3/2}.
\end{align}
Assuming an isotropic stress state, we can then rewrite equation \eqref{eqn:kinetic-no-q} as
\begin{align}
    (3/2)\rho \dot{T} = 4 \rho \tilde{G} \tilde{F} T \tr{\symvelgrad} - \frac{12}{\sqrt{\pi}} \frac{\rho \tilde{G}}{d} (1 - \epsilon^2) T^{3/2}.
\end{align}

This is an ordinary differential equation of the form
\begin{align}
    \dot{T} = A T - B T^{3/2}
\end{align}
where
\begin{align}
    A &= (8/3) \tilde{G} \tilde{F} \tr{\symvelgrad} \\
    B &= \frac{8}{\sqrt{\pi}} \frac{\tilde{G}}{d} (1 - \epsilon^2).
\end{align}

Note first that $T = 0$ is a solution to this differential equation (for all times).
Otherwise, the temperature decreases or remains at the current value when
\begin{align}
    \frac{3}{d \sqrt{\pi}} (1 - \epsilon^2) \sqrt{T} \ge \tilde{F} \tr{\symvelgrad}.
\end{align}

We may also integrate the general form and rearrange to obtain the expression for the granular temperature as a function of time, given by
\begin{align}
    T(t) = \frac{A^2 \exp \left(At + AC\right)}{\left(1+B \exp \left(\frac{At + AC}{2}\right)\right)^2},
\end{align}
where $C$ is a constant set by the initial conditions.
We have discarded a nonphysical solution where the temperature contains a singularity at a finite positive time.
In the limit of large times, even with a constant volumetric compression rate we note that the temperature is limited to $\frac{A^2}{B^2}$.
Thus, as long as the material initially begins with a zero temperature, experiences a short-duration compression event, or experiences a long-duration compression event where $A/B$ is small, the kinetic stresses remain small.

We then take the upper bound for this temperature as given by
\begin{align}
T_c &= \frac{A^2}{B^2} = \frac{\pi}{144(\lambda - 1)^2} \left( 1 + \frac{1}{4 \tilde{G} \lambda} \right)^2 \dot{\epsilon}_v^2 d^2
\end{align}
where we have written $\tr \symvelgrad$ as $\dot{\epsilon}_v$.

A similar analysis can be done for the volumetric term as the shear term. The pressure is given by
\begin{align}
p(\Phi) = \frac{\pi \rho_s \Phi (4 \tilde{G} \lambda + 1) }{144(\lambda - 1)^2} \left( 1 + \frac{1}{4 \tilde{G} \lambda} \right)^2 \dot{\epsilon}_v^2 d^2.
\end{align}
The inverse relation can be used to obtain $\Phi(p, \dot{\epsilon}_v)$, however unlike in the shear case the equation is quartic in $\Phi$ and is cumbersome to manipulate algebraically.

As before, this expression can be combined with the elastic contribution given by
\begin{align}
\frac{v_e}{m} = \frac{-p}{\rho_s K_c}
\end{align}
in a manner analogous to the shear case to obtain a pressure $p$ at a given $\Phi$ for given values of $\dot{\gamma}$ and $\dot{\epsilon}$.
Since errors occur in our approximation when $p(\Phi_c)$ is large, this value may be calculated at each instant to obtain an estimate of the error incurred in taking the pressure as zero below the critical value of $\rho_c = \rho_s \Phi_c$.
The calculation may be done in a fully coupled manner, where the inverse packing fraction contributions come from elasticity, a shear flowing part, and a volumetric flowing part, or the calculation can be done for a reduced system (with the elastic part combined with only the shear term or only the volumetric term).
For a conservative apriori estimation we can take the strain rate as $u_c / d$, where $u_c$ is the maximum velocity in the system.

Although we do not track the evolution of flowing density when at a non-zero pressure, we may wish to reconstruct the density variation in the steady state.
A typical post-processing method (e.g. mentioned in \citet{pouliquen09}) to compute the density variation is to use the inertial number $I$ which was obtained in simulation and use the $\phi(I)$ relationship to recover $\phi$.
This does not impact the mass balance as long as the modified $\rho$ statisfies
\begin{align}
    0 = - \mathbf{v} \cdot \nabla \rho - \rho \nabla \cdot \mathbf{v},
\end{align}
which is the Eulerian form for mass conservation in steady state conditions.
We may estimate the error incurred in performing this conversion by computing how closely the steady-state fields satisfy this equation.

\section{Verification of Constitutive Integration Scheme}
\label{app:verification}
Testing our stress update procedure against an exact solution given some simple, time-dependent velocity gradient allows us to determine error in the constitutive integration.
While dense, we know that the stress state evolves according to equation  \eqref{eqn:constitutive-cases}.
If we write this in component form and combine terms, we can find (nonlinear) ODEs for the stress components given a velocity gradient.
Even in the simplest cases, analytical solutions to these ODEs are difficult to obtain; instead, we will consider a high-order numerical solution as a reference solution.
We used the fourth-order Runge-Kutta integrator in the SciPy package from \citet{jones01}, which implements the algorithm described in \citet{dormand80}.
We use the parameters in table \ref{tbl:material-params}, with the exception of the Young's modulus, which is reduced to 10000 Pa to lessen the time step size restriction. The applied velocity gradient is given over the time interval $t = 0$ s to $t = 1.25$ s, and the components are chosen as
\begin{align}
    \velgrad(t) = \velgrad_{\mathrm{shear}} + (H(t-0.25) - H(t-0.5))\velgrad_{\mathrm{com}} + (H(t-0.75) - H(t-1))\velgrad_{\mathrm{ext}}
\end{align}
where
\begin{align}
\velgrad_{\mathrm{shear}} = 
\begin{bmatrix}
    0 & 0.05 & 0\\
    0 & 0 & 0 \\
    0 & 0 & 0
\end{bmatrix}, \quad \velgrad_{\mathrm{com}} = (|8t - 3| - 1)\begin{bmatrix} 1 & 0 & 0 \\ 0 & 1 & 0 \\ 0 & 0 & 0 \end{bmatrix}, \quad \velgrad_{\mathrm{ext}}= (1 - |8t - 7|)\begin{bmatrix} 1 & 0 & 0 \\ 0 & 1 & 0 \\ 0 & 0 & 0 \end{bmatrix},
\end{align}
and $H(t)$ is the Heaviside step function. This represents steady planar shearing combined with time-varying compression.
\begin{figure}
\centering
\includegraphics[scale=1.0]{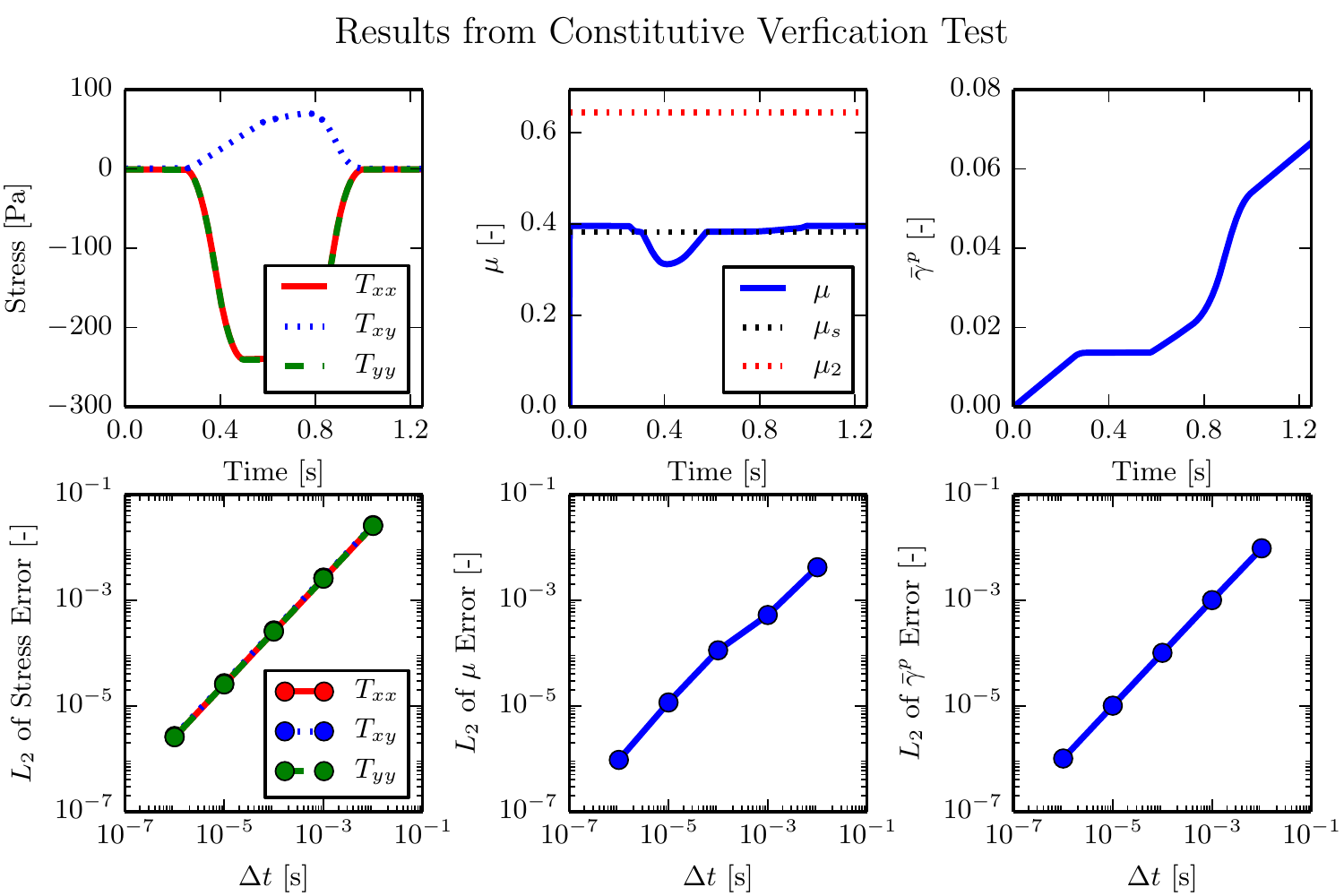}
\caption{The results of our constitutive integration algorithm (top row) and the comparison to the fourth order Runga Kutta scheme (bottom row). The input is the velocity gradient, $\velgrad$, and the outputs are the stress and the plastic strain-rate tensor. The results of our constitutive integration closely track those of the fourth order Runge-Kutta scheme and converge with first order error in the time step as expected.}
\label{fig:compare-integ}
\end{figure}
The results of our constitutive integration algorithm are plotted in figure \ref{fig:compare-integ} (top row), which are in good agreement with the results from the fourth order Runge-Kutta integrator (differences in bottom row). Note that our test is different than the method of manufactured solutions (MMS) as explained in \citet{kamojjala13}, which exercises the entire code path and performs a comprehensive test of an implementation itself. The verification presented here only extends to the numerical integration of the stress as a function of velocity gradient.


}

\end{document}